%% file: main.tex
\documentclass[12pt,3p,doublespacing]{elsarticle}
\journal{.}
\usepackage{times}
\usepackage{libertine}
\emergencystretch=\maxdimen
\usepackage{sectsty}
\sectionfont{\color{blue}\large\MakeUppercase}
\subsectionfont{\color{blue}}
\biboptions{sort&compress}
\usepackage[colorlinks]{hyperref}
\AtBeginDocument{%
 	\hypersetup{
	plainpages = false, 
	bookmarksopen = true,
	bookmarksnumbered = true,
	breaklinks = true,
	linktocpage,
	colorlinks = true,
	linkcolor = purple,
	urlcolor  = black,
	citecolor = Magenta,
}}
\usepackage{multicol}
\usepackage{multirow}
\usepackage{imakeidx}
\makeindex
\usepackage{nomencl}
\makenomenclature
\usepackage[xindy]{glossaries}
\makeglossaries
\usepackage{graphics}
\usepackage{graphicx}
\usepackage{epsf,psfrag}
\usepackage{epstopdf}
\usepackage[para]{threeparttable}
\usepackage{subfigure}
\usepackage{epsfig}
\usepackage{pdflscape}
\usepackage{rotating}
\usepackage{lscape}
\usepackage{float}
\usepackage{stfloats}
\usepackage{textgreek}
\usepackage{longtable}
\usepackage{lscape}
\usepackage{capt-of}
\usepackage{tablefootnote}
\usepackage{footnote}
\usepackage{caption}
\usepackage[autopunct=true]{csquotes}
\usepackage{rotating}
\usepackage[normalem]{ulem}
\usepackage{resizegather}
\usepackage{siunitx}
\usepackage[nodisplayskipstretch]{setspace}
\usepackage{color,soul}
\usepackage[usenames,dvipsnames]{xcolor}
\usepackage{amsfonts,amsmath,amssymb,gensymb}
\usepackage{latexsym,array}%,subeqn}
\usepackage{textcomp}

\newcommand{\RomanNumeralCaps}[1]
\linenumbers
\makesavenoteenv{tabular}
\makesavenoteenv{table}
\usepackage[left,displaymath, mathlines]{lineno}
\DeclareMathAlphabet{\mathpzc}{OT1}{pzc}{m}{it}
\def\fig{Fig.~}
\def\figs{Figs.~}
\def\eqn{Eq.~}
\def\eqns{Eqs.~}
\def\tab{Table~}

\newcommand{\myvec}[1]{\mathbf{#1}}     
\def\tsc#1{\csdef{#1}{\textsc{\lowercase{#1}}\xspace}}
\tsc{WGM}
\tsc{QE}
\tsc{EP}
\tsc{PMS}
\tsc{BEC}
\tsc{DE}
\usepackage{easyReview} %\setreviewsoff
\newcommand{\removeEq}[1]{\ifistoreview{\@\expandafter\removeColor{#1}\hspace{-0.6em}} \else {}\fi}
\newcommand{\ttxt}[1]{\textbf{\color{blue}\large\MakeUppercase{#1}}}     
\graphicspath{{../}{Figures/}{Figures/Validation/}}
\DeclareGraphicsExtensions{.png, .PNG,.pdf,.PDF,.jpg,.JPG,.eps,.EPS}
\begin{document}

%--------------------------------------------
%\input{reviewer-1.tex}\newpage
%\setcounter{page}{1}
%
%--------------------------------------------
%\input{highlights.tex}\newpage
%--------------------------------------------
\setcounter{page}{1}
\begin{frontmatter} % for elsevier jrnls
% 
%\openup -0.5em % 1em for double spacing
\title{\ttxt{CFD analysis of the influence of contraction size on electroviscous flow through the slit-type non-uniform microfluidic device}}
\author[addr]{Jitendra {Dhakar}}
%\ead{jdhakar@ch.iitr.ac.in}
\author[addr]{Ram Prakash {Bharti}\corref{coradd}}
\ead{rpbharti@iitr.ac.in}
\address[addr]{Complex Fluid Dynamics and Microfluidics (CFDM) Lab, Department of Chemical Engineering, Indian Institute of Technology Roorkee, Roorkee - 247667, Uttarakhand, INDIA}
%
%\address[labela]{Department of Chemical Engineering, Indian Institute of Technology Roorkee, Roorkee - 247667, Uttarakhand, INDIA}
%
\cortext[coradd]{\textit{Corresponding author. }}
%
%%%%%%%%%%%%%%%%%%%%%%%%%%%%%%%%%%%%%%%%%%%%%%%%%%%%%%%%%%%%%%%%%%%%%%%%%%%%%%%%%%%%%%
\begin{abstract}
	\fontsize{12}{22pt}\selectfont
	%----Text of abstract
	\noindent 
	The electroviscous effects are relevant in controlling and manipulating the fluid, thermal, and mass transport microfluidic processes. The existing research has mainly focused on the fixed contraction ratio ($d_\text{c}$, i.e., the area ratio of contraction to expansion) concerning the widely used contraction-expansion geometrical arrangement. This study has explored the influence of the contraction ratio ($d_\text{c}$) on the electroviscous flow of electrolyte liquids through the charged non-uniform microfluidic device. The numerical solution of the mathematical model (Poisson's, Nernst-Planck, and Navier-Stokes) using a finite element method (FEM) yields the local flow fields. In general, the contraction ratio significantly affects the hydrodynamic characteristics of microfluidic devices. The total electrical potential and pressure drop maximally increase by {1785\% and 2300\%}, respectively, with an overall contraction ratio ($0.25\le d_\text{c}\le 1$). Further, an electroviscous correction factor ($Y$, i.e., the ratio of apparent to physical viscosity) maximally enhances by 11.24\% (at $K=8$, $S=16$ for  $0.25\le d_\text{c}\le 1$), 31.80\% (at $S=16$, $d_\text{c}=0.75$ for $2\le K\le 20$), 22.89\% (at $K=2$, $d_\text{c}=0.5$ for $4\le S\le 16$), and 46.99\% (at $K=2$, $d_\text{c}=0.75$ for $0\le S \le 16$). The present numerical results may provide valuable guidelines for the performance optimization and design of reliable and essential microfluidic devices.
\end{abstract}
%%%%%%%%%%%%%%%%%%%%%
\begin{keyword}
	\fontsize{12}{22pt}\selectfont
	%----keywords here, in the form: keyword \sep keyword
	Electrolyte liquid\sep Microfluidic device\sep Contraction ratio\sep Pressure-driven flow \sep Pressure drop \sep Electroviscous effect
	% \sep partially heated wall \sep square cavity
	%----PACS codes here, in the form: \PACS code \sep code
\end{keyword}
\end{frontmatter}
%%%%%%%%%%%%%%%%%%%%%%%%%%%%%%%%%%%%%%
%
%\fontsize{12}{14pt}\selectfont
%\linespread{1.6}
%\doublespacing
%\openup 1em % 1em for double spacing
\setstretch{2} 
%
%%%%%%%%%%%%%%%%%%%%%%%%%%%%%%%%%%%%%%%%%%%%%%%%%%%%%%%%%%%%%%%%%%%%%
\section{Introduction}
\label{sec:intro}
%%%%%%%%%%%%%%%%%%%%%%%%%%%%%%%%%%%%%%%%%%%%%%%%%%%%%%%%%%%%%%%%%%%%%
%
\noindent 
Due to the rapid development of Lab-on-chip (LOC) devices, microscale channels are widely used in various engineering and biomedical applications, such as DNA analysis \citep{foudeh2012microfluidic,bruijns2016microfluidic,li2021microfluidic}, drug delivery \citep{vladisavljevic2013industrial,nguyen2013design}, medical diagnosis sensors \citep{figeys2000lab,wu2018lab}, micro heat exchangers \citep{xue2001performance,stogiannis2015efficacy,bahiraei2017efficacy,pan2019numerical}, and electronic chip cooling \citep{imran2018numerical,abdulqadur2019performance,tan2019heat,zhuang2020optimization}. 
In particular, LOC devices are a very efficient way to transport, manipulate and control the flow of electrolyte liquids. Therefore, understanding the fluid dynamics of microscale geometries is essential due to their inherent characteristics, such as the effects of the surface forces (i.e., electroviscous, electromagnetic, surface tension forces, etc.), which are either absent or non-dominant in conventional macroscale geometries. Among others, electroviscous (EV) effects are quite common in electrolyte flow through microdevices due to the induced electric field intensely manipulating the fluid dynamics.
\\\noindent 
Electroviscous (EV) effects arise due to an interaction of electrolyte liquid with the charged walls of the microfluidic device in the pressure-driven flow (PDF). The charged surfaces attract the counter-ions (and repeal co-ions) and, thus, form an `electrical double layer' (EDL), consisting of compact and diffuse layers, in close vicinity of the surface \citep{hunter2013zeta,li2001electro}. The convection flow of excess ions in EDL develops a `streaming current' ($I_\text{s}$), and the corresponding potential is termed the `streaming potential' (\fig\ref{fig:1}). Subsequently, a `conduction current' ($I_\text{c}$), generates the opposite of the pressure-driven flow (PDF) and contributes as an additional hydrodynamic resistance. It results in the retarding of the net flow of liquid in the direction of the primary PDF due to the commonly called `electroviscous effect' \citep{hunter2013zeta,davidson2007electroviscous,pimenta2020viscous,dhakar2022electroviscous}.
%
%\begin{comment}
%	In this study, electroviscous (EV) effects are analyzed in the microfluidic devices that arise due to contact of pressure-driven flow (PDF) of electrolyte liquid with the charged walls of microchannel. Charged surfaces of device attract counter-ions and repeal co-ions; thus, redistribution of ions close to the surface formed an `electrical double layer' (EDL) \citep{hunter2013zeta,li2001electro}. The convective flow of excess ions in the EDL develops a `streaming current' and potential related to this current is known as `streaming potential' (refer \fig\ref{fig:1}). It generates a `conduction current' opposite to the PDF that retards the net flow of liquid by adding the additional hydrodynamics forces (refer \fig\ref{fig:1}). This results reduced net flow rate of liquid in the direction of primary PDF and it is commonly called as the `electroviscous effect' \citep{hunter2013zeta,davidson2007electroviscous,pimenta2020viscous,dhakar2022electroviscous}. 
%\end{comment}
%
\\\noindent 
Further, non-uniform geometries (e.g.,  contraction, contraction-expansion,  converging-diverging, X- Y-, T-junctions, and other branching) are commonly used components in the fabrication of the microfluidic devices \citep{davidson2007electroviscous}. In contraction-expansion geometries, additional (i.e., excess) pressure drop resulting from the sudden contraction/expansion enhances the overall pressure drop in the microfluidic devices \citep{Sisavath2002,bharti2008steady,bharti2009electroviscous,davidson2007electroviscous,davidson2008electroviscous,davidson2010electroviscous,dhakar2022slip,dhakar2022electroviscous,dhakar2023cfd}. 
%The contraction ratio ($d_{\text{c}}={A_{\text{c}}}/{A}$, where, $A_\text{c}$ and $A$ are the cross-sectional area of contraction and uniform expansion, respectively), plays a vital role in the hydrodynamics of microfluidic systems
The contraction ratio ($d_{\text{c}}$) plays a vital role in the hydrodynamics of microfluidic systems and, in general,  is defined as follows. 
\begin{gather}
	d_{\text{c}}=\frac{A_{\text{c}}}{A}, \quad 0< d_\text{c}\le 1
	\label{eq:12}
\end{gather}
where, $A_\text{c}$ and $A$ are the cross-sectional area of contraction and uniform expansion, respectively.
\\
\noindent 
The existing literature concerning electroviscous flow (EVF) in microfluidic devices has been reviewed in our recent studies \citep{dhakar2022slip,dhakar2022electroviscous,dhakar2023cfd}, and thus not repeated here.
A voluminous literature has explored various aspects of the electroviscous (EV) effects in the pressure-driven flow (PDF) through charged uniform ($d_\text{c}=1$) microfluidic devices like slit \citep{burgreen1964electrokinetic,mala1997flow,mala1997heat,chun2003electrokinetic,ren2004electroviscous,chen2004developing,joly2006liquid,xuan2008streaming,wang2010flow,jamaati2010pressure,zhao2011competition,tan2014combined,jing2015electroviscous,matin2016electrokinetic,jing2017non,matin2017electroviscous,kim2018analysis,sailaja2019electroviscous,mo2019electroviscous,li2021combined,li2022electroviscous,banerjee2022analysis,Xing2023}, rectangular \citep{yang1998modeling,li2001electro,ren2001electro}, elliptical \citep{hsu2002electrokinetic}, and cylinder \citep{rice1965electrokinetic,levine1975theory,bowen1995electroviscous,brutin2005modeling,bharti2009electroviscous,jing2016electroviscous}.
On the other hand, few studies have understood the electroviscous effects by measuring the pressure drop in the non-uniform (contraction-expansion) geometries for a fixed contraction size ($d_\text{c}=0.25$, $L_\text{c}=L_\text{u}=L_\text{d}$) for contraction-expansion slit \citep{davidson2007electroviscous,berry2011effect,dhakar2022slip,dhakar2022electroviscous,dhakar2023cfd}, cylinder \citep{bharti2008steady,davidson2010electroviscous}, and rectangular \citep{davidson2008electroviscous} geometries. 
Broadly, these studies \citep{bharti2008steady,bharti2009electroviscous,davidson2007electroviscous,davidson2008electroviscous,davidson2010electroviscous,dhakar2022slip,dhakar2022electroviscous,dhakar2023cfd} have shown that surface charge density ($4\le S \text{ or } S_\text{t}\le 16$), surface charge asymmetry ($S_\text{r}=S_\text{b}/S_\text{t}$, $0\le S_\text{r}\le 2$), inverse Debye length ($2\le K\le 20$), and slip length ($0\le B_0\le 0.20$) have remarkably affected the flow characteristics in microfluidic devices at a fixed volumetric flow rate ($Q$). Subsequently, they \citep{bharti2008steady,bharti2009electroviscous,davidson2007electroviscous,davidson2008electroviscous,davidson2010electroviscous,dhakar2022slip,dhakar2022electroviscous,dhakar2023cfd} have presented predictive models for pressure drop, and electroviscous correction factor ($Y=\Delta P/\Delta P_0$, where subscript `0' indicates an absence of electrical influences)  by accounting the pressure drop in uniform planer channel/pipe by Poiseuille flow, and excess pressure drop due to sudden contraction/expansion under creeping flow. These analytical models maximally overestimated the pressure drop $\pm$5\% compared to their numerical results.  
\\\noindent 
In summary, the literature comprises knowledge of the electroviscous effects in liquid flow through uniform ($d_\text{c}=1$) or non-uniform ($d_\text{c}=0.25$) geometries. While the non-uniform geometries have shown significant influence on the pressure drop and electroviscous effects, the impact of the contraction ratio variation ($d_\text{c}$)  on the electroviscous flow through charged non-uniform microfluidic devices has been unexplored in the literature, to the best of our knowledge.
%
%\\\noindent 
Hence, the present work aims to numerically investigate the influence of the contraction ratio ($d_\text{c}$) on the electrolyte liquid flow through charged non-uniform slit microfluidic devices. A finite element method (FEM) is used to solve the governing equations such as Poisson's, Nernst-Planck, and Navier-Stokes equations to obtain the flow fields like total potential ($U$), {excess ionic charge} ($n_\pm$), induced electric field strength ($E_\text{x}$), velocity ($\myvec{V}$), and pressure ($P$), for the broader ranges of dimensionless parameters ($2\le K\le 20$, $4\le S\le 16$, $0.25\le d_\text{c}\le 1$). Finally, a pseudo-analytical model is developed to estimate the pressure drop ($\Delta P$) and electroviscous correction factor ($Y$) for broader ranges of governing parameters ($K$, $S$, $d_\text{c}$).
%%%%%%%%%%%%%%%%%%%%%%%%%%%%%%%%%%%%%
%
\section{Problem statement}
\begin{figure}[!b]
	\centering
	\includegraphics[width=1\linewidth]{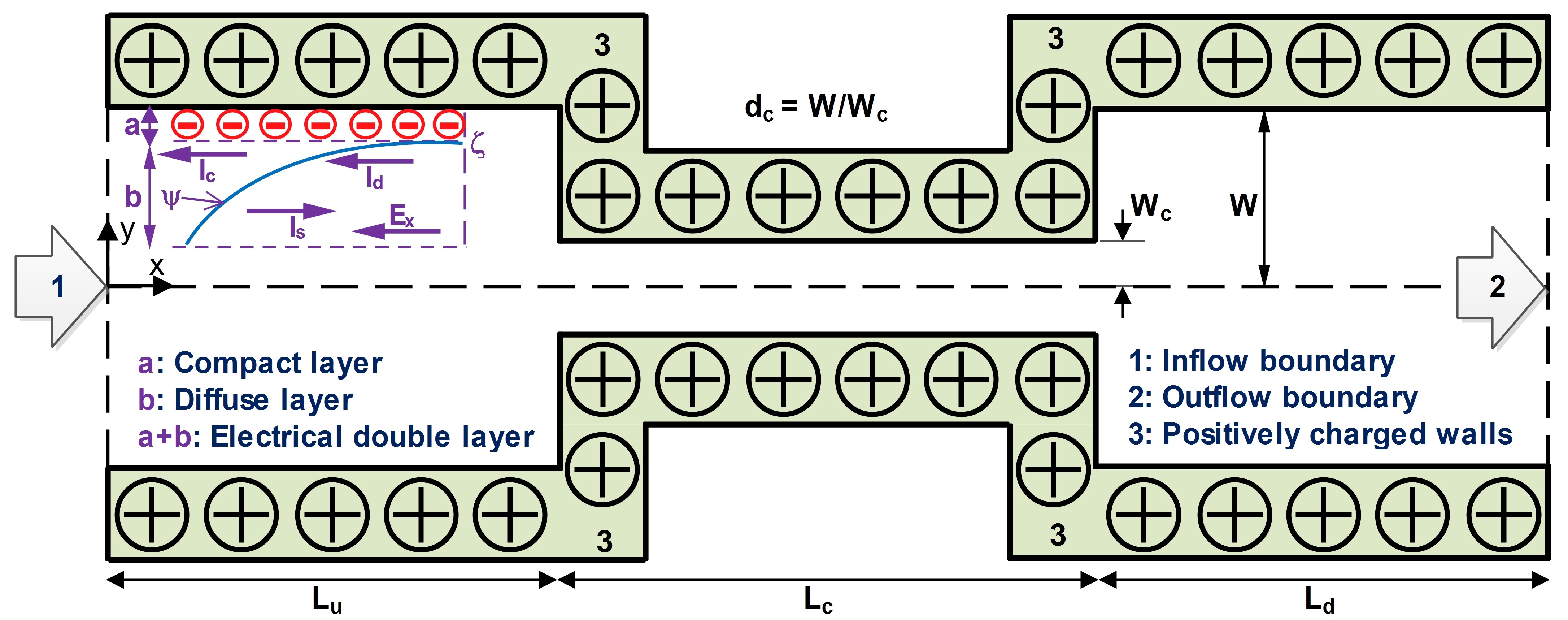}
	\caption{Schematics of pressure-driven flow of liquid electrolyte through charged non-uniform slit microfluidic device.}
	\label{fig:1}
\end{figure}

%%%%%%%%%%%%%%%
\noindent \fig\ref{fig:1} displays the pressure-driven flow of electrolyte liquid through a non-uniform (i.e., contraction-expansion) slit microfluidic device. 
%
%, accounting the influence of electroviscous and contraction variation effects. 
%
% %
The two-dimensional (2-D) flow is laminar and fully-developed with an average inflow velocity of $\overline{V}$ m/s, and the channel walls are uniformly charged with positive charge density ($\sigma$, C/m$^2$).  
The contraction section is situated between the upstream and downstream sections in the microchannel. The geometrical characteristics (length, width in $\mu$m) of upstream, contraction, and downstream sections of the microfluidic device are ($L_\text{u}$, $2W$), ($L_\text{c}$, $2W_\text{c}$), and ($L_\text{d}$, $2W$), respectively; thus, the total length of the microfluidic device is $L~(=L_\text{u}+L_\text{c}+L_\text{d})$, and the contraction ratio is $d_\text{c}$ (\eqn\ref{eq:12}). The subscripts `u', `c', and `d' represent the upstream, contraction, and downstream.
%
%Contraction section is placed between upstream and downstream sections in the microchannel. The geometrical dimensions (length, width) of upstream, contraction, and downstream sections are ($L_\text{u}$, $2W$), ($L_\text{c}$, $2W_\text{c}$), and ($L_\text{d}$, $2W$), respectively. The $d_\text{c}$ (\eqn\ref{eq:12}) is the contraction ratio and $L(=L_\text{u}+L_\text{c}+L_\text{d})$ is the total length of microfluidic device. 
%
\\\noindent 
The liquid electrolyte is Newtonian (viscosity, $\mu$ Pa.s), incompressible (density, $\rho$ kg/m$^3$), symmetric (i.e., equal cations and anions, $n_{+}=n_{-}$ or 1:1) with equal ionic valences ($z_{+}= - z_{-}=z$) and diffusivity ($\mathcal{D}_{+}=\mathcal{D}_{-}=\mathcal{D}$, m$^2$/s). The bulk (i.e., geometric mean) ionic concentration  \citep{harvie2012microfluidic,davidson2016numerical} of liquid is $n_0$ m$^{-3}$. The dielectric constant of liquid ($\varepsilon_{\text{r}}$) is spatially constant, and that of the wall is considerably small than liquid ($\varepsilon_{\text{r,w}}<<\varepsilon_{\text{r}}$).
%
%Liquid is assumed as symmetric ($1$:$1$) electrolyte having equal ionic valences ($z_\text{+}=z_\text{-}=z$) and equal diffusivity ($D_\text{+}=D_\text{-}=D$).  The bulk (i.e., geometric mean) ionic concentration is $n_0$ \citep{harvie2012microfluidic,davidson2016numerical,dhakar2022electroviscous}. 
%Surface of the microfluidic device is considered to have uniform positive charge with $\sigma$ surface charge density on the walls of device. The liquid is considered as Newtonian and incompressible, i.e., viscosity ($\mu$), density ($\rho$), and dielectric constant of liquid ($\varepsilon_{\text{r}}$) are spatially constant. The dielectric constant of wall is considered to be very small than liquid ($\varepsilon_{\text{r,w}}<<\varepsilon_{\text{r}}$).
%
\subsection{Governing Equations and Boundary Conditions}
\noindent 
The above-stated physical problem of electrolyte flow through the charged microfluidic device is, mathematically, governed by Poisson's, Nernst-Planck (N-P), and Navier-Stokes (N-S) equations. The dimensional form of the governing differential equations and boundary conditions are presented in the literature \citep{dhakar2022electroviscous}, and thus, not repeated here. The mathematical equations  \citep{dhakar2022electroviscous} are non-dimensionalized using the scaling factors such as $\bar{V}$, $W$, $n_0$, $W/\bar{V}$, $U_c~(=k_\text{B}T/ze)$ for velocity, length, the number density of ions, time, and electrical potential, respectively. The dimensionless groups (Reynolds number $Re$, Schmidt number $\mathit{Sc}$, Peclet number $Pe$, inverse Debye length $K$, liquid parameter $\beta$, surface charge density $S$) obtained from scaling analysis of governing equations and boundary conditions are expressed as follows.
%\noindent The electroviscous (EV) flow in charged slit microfluidic device is numerically solved by the Poisson's (\eqn\ref{eq:2}), Nernst-Planck (\eqn\ref{eq:3}), Navier-Stokes (\eqn\ref{eq:4}), and continuity (\eqn\ref{eq:5}) equations as follows. The governing equations and boundary conditions in the dimensional form are explained and presented elsewhere \citep[refer \eqns A.1 to A.16 in][]{dhakar2022electroviscous}. Scaling factors such as $\bar{V}$, $W$, $n_0$, $W/\bar{V}$, $U_c(=k_\text{B}T/ze)$ for velocity, length, number density of bulk ions, time, and electrical potential, respectively are used for non-dimensionlization of these governing equations.  The dimensionless groups obtained from scaling analysis are shown as follows.
%
%
\begin{gather}
	Re=\frac{\rho\bar{V}W}{\mu}; \quad
	\mathit{Sc}=\frac{\mu}{\rho \mathcal{D}}; \quad
	Pe =Re\times\mathit{Sc}; \quad
	\beta=\frac{\rho\varepsilon_{\text{0}}\varepsilon_{\text{r}}U_c^2}{2\mu^2}; \quad 
	K^2=\frac{2W^2zen_{\text{0}}}{\varepsilon_{\text{0}}\varepsilon_{\text{r}} U_c}; \quad
	S=\frac{\sigma W}{\varepsilon_{\text{0}}\varepsilon_\text{r} U_c}
	\label{eq:1}
\end{gather}
%
%where $K$ and $\beta$ are the inverse Debye length and liquid parameter, respectively. The $Pe$, $Re$, and $\mathit{Sc}$ are the Peclet, Reynolds, and Schmidt numbers, respectively.
%
%The above-state present physical problem can be mathematically expressed by the following set of dimensionless governing equations:
where, the permittivity of free space ($\varepsilon_{\text{0}}$, F/m),  Boltzmann constant ($k_\text{B}=1.380649 \times 10^{-23}$ J/K), temperature ($T$, K), elementary charge ($e=1.602176634 \times 10^{-19}$ C),  respectively. 
Hereafter, all the variables, fields, and equations are expressed in the dimensionless form. 
\\\noindent 
The Poisson's (\eqn\ref{eq:2}), Nernst-Planck (\eqn\ref{eq:3}), and Navier-Stokes (\eqns\ref{eq:4} - \ref{eq:5}) equations, in dimensionless form, are written as follows.
\begin{gather}
	\nabla^2U=-\frac{1}{2}K^2n^\ast 
	%	\qquad \text{where} \qquad n^\ast=(n_{\text{+}}-n_{\text{-}})
	\label{eq:2}
	\\
	\left[\frac{\partial n_{\text{j}}}{\partial t}+\nabla\cdot(\myvec{V}n_{\text{j}})\right]=Pe^{-1}\left[\nabla^2n_{\text{j}}\pm\nabla\cdot(n_{\text{j}}\nabla U)\right]
	\label{eq:3}
	\\
	\left[\frac{\partial \mathbf{V}}{\partial t}+\nabla\cdot(\myvec{V}\myvec{V})\right]=-\nabla P+Re^{-1}\nabla \cdot\left[\nabla\myvec{V}+(\nabla\myvec{V})^T\right]-\underbrace{(\beta K^2Re^{-2})n^\ast\nabla U}_{\myvec{F}_{\text{e}}}
	\label{eq:4}
	\\
	\nabla\cdot\myvec{V}=0 \label{eq:5}
\end{gather}
where $\myvec{V}$, $P$, $n_\text{j}$, $n^\ast(=n_{\text{+}}-n_{\text{-}})$, and $\myvec{F}_{\text{e}}$ are velocity vector, pressure, number density of $j^{th}$ ion, excess charge, and electrical body force, respectively. 
%
%\\\noindent 
The total electrical potential ($U$) is the sum of EDL potential ($\psi$) and streaming potential ($\phi = \phi_{0} - xE_\text{x}$) in the electroviscous flows (EVFs), i.e.,
\begin{gather}
	U(x,y)=\phi(x) + \psi(y)
	\label{eq:1a}
\end{gather}
where $\phi_{0}$ and $E_\text{x}$ are the reference potential at the inlet ($x=0$), and axially induced electric field strength, respectively. Notably, two potentials ($\phi$ and $\psi$) act independently as the field directions throughout remain the same for the uniform ($d_\text{c}=1$) geometries. They, however, are strongly coupled for the non-uniform ($d_\text{c}\neq1$)  geometries, and the total potential ($U$) is thus computed and analyzed \citep{davidson2007electroviscous,bharti2008steady,bharti2009electroviscous,dhakar2022electroviscous,dhakar2023cfd}.   
\\\noindent 
The governing equations (\eqns\ref{eq:2} to \ref{eq:5}) are subjected to the  following relevant dimensionless boundary conditions:
\\\noindent 
(i) The fully developed velocity  ($\myvec{V}$) and ionic concentration ($n_\pm$) fields obtained numerically/analytically \citep{davidson2007electroviscous,dhakar2022electroviscous}, and uniform axial potential gradient  ($\partial_x U$) are imposed at the inlet ($x=0$) boundary as follows.
%Velocity ($\myvec{V}$) and ionic concentration ($n_\pm$) fields are imposed at inlet ($x=0$) boundary of device as follows.
%
\begin{gather}
	%\begin{gather}
	V_{\text{x}}=V_{0}(y), \qquad
	V_{\text{y}}=0, \qquad
	%n_{{+}}=e^{-\psi(y)}, \qquad 
	%n_{{-}}=e^{\psi(y)}
	n_{{\pm}}=\exp[{\mp\psi_0(y)}],  \qquad
	%	n_{{-}}=\exp[{\psi(y)}]
	\frac{\partial U}{\partial x} = \text{C}_1
	\label{eq:6} 
\end{gather}
where $V_{0}$ and $\psi_0$ are the fully developed velocity and EDL potential fields for uniform slit.  The constant ($\text{C}_1$) is obtained by satisfying the ``current continuity condition'' (i.e., net current, $I_\text{net}=\nabla\cdot I=0$, \eqn\ref{eq:7}) expressed \citep{davidson2007electroviscous,bharti2008steady,bharti2009electroviscous,dhakar2022electroviscous} as follows.
\begin{gather}
	I_{\text{net}} = \underbrace{\int_{-1}^{1} {n^\ast\myvec{V}} dy} _{I_{\text{s}}} + \underbrace{\int_{-1}^{1} -{Pe^{-1}\left[\frac{\partial n_{\text{+}}}{\partial x}-\frac{\partial n_{\text{-}}}{\partial x}\right]} dy}_{I_{\text{d}}} + \underbrace{\int_{-1}^{1} -{Pe^{-1}\left[(n_{\text{+}}+n_{\text{-}})\frac{\partial U}{\partial x}\right]} dy}_{I_{\text{c}}} =0
	\label{eq:7}
\end{gather}
where $I_{\text{c}}$, $I_{\text{d}}$, and $I_{\text{s}}$ are the conduction, diffusion and streaming currents, respectively. The diffusion current is zero ($I_\text{d}=0$) at the steady-state condition.
\\\noindent 
(ii) Both flow and ionic concentration fields are allowed to be fully developed, and uniform axial potential gradient  ($\partial_x U$) is maintained at the outlet ($x=L$) boundary open to ambient as follows.
\begin{gather}
	%	\frac{\partial \myvec{V}}{\partial \myvec{n}_{\text{b}}} = 0,
	\frac{\partial \myvec{V}}{\partial x} = 0,
	\qquad
	P =0, 
	\qquad
	%	\frac{\partial n_{\text{j}}}{\partial \myvec{n}_{\text{b}}} = 0,
	\frac{\partial n_{\text{j}}}{\partial x} = 0,
	\qquad
	\frac{\partial U}{\partial x} = \text{C}_2
	\label{eq:8}
\end{gather}
The constant ($\text{C}_2$) is obtained by satisfying \eqn(\ref{eq:7}) at the outlet ($x=L$) boundary.
\\\noindent 
(iii) No-slip, solid, impermeable walls of the device are uniformly charged, i.e., 
\begin{gather}
	\myvec{V} = 0,
	%	V_{\myvec{n}_{\text{b}}} =0,
	%\qquad\mbox{and}
	%	\qquad 
	%	V_{\myvec{t}_{\text{b}}} =0,
	\qquad
	\myvec{f}_{\text{j}}\cdot \myvec{n}_{\text{b}}=0,\qquad
	(\nabla U\cdot\myvec{n}_{b}) = S
	\label{eq:9}
\end{gather}
where $\myvec{f}_{\text{j}}$ is the flux density of ions described by Einstein relation \citep{dhakar2022electroviscous,ZLi2023}, and $\myvec{n}_{b}$ is unit vector normal to the boundary. 
%\\\noindent \eqns(\ref{eq:6}) and (\ref{eq:7}) are imposed at the inlet ($x=0$) of microfluidic device. \eqns(\ref{eq:7}) and (\ref{eq:8}) are imposed at the outlet ($x=L$) of microfluidic device. \eqns(\ref{eq:9}) to (\ref{eq:11}) are imposed at the walls of microfluidic device.
%
\\\noindent 
The coupled mathematical model (\eqns\ref{eq:2} to \ref{eq:9}) has been numerically solved to obtain the flow fields, i.e., total potential ($U$), ionic concentration ($n_\pm$), velocity ($\myvec{V}$), pressure ($P$), excess charge ($n^*$), and induced electric field strength ($d_\text{x}$) fields, in the non-uniform microfluidic device for wide ranges of dimensionless parameters ($2\le K\le 20$, $4\le S\le 16$, $0.25\le d_\text{c}\le 1$).
%A finite element method (FEM) is used to solve the mathematical model consisting of dimensionless governing equation (\eqns\ref{eq:2} to \ref{eq:5}) with relevant boundary conditions (\eqns\ref{eq:6} to \ref{eq:9}) to obtain the flow fields i.e., total potential ($U$), ionic concentration ($n_\pm$), velocity ($\myvec{V}$), and pressure ($P$) fields for wide ranges of dimensionless parameters ($2\le K\le 20$, $4\le S\le 16$, $0.25\le d_\text{c}\le 1$) in the microfluidic device.
%
\\\noindent 
At this point, it is appropriate to briefly discuss the important quantities and definitions used in the subsequent discussion of the new results. For instance, 
in order to explore the influence of the contraction ratio ($d_\text{c}$), the hydrodynamic quantities (say, $\lambda$) have been normalized, under otherwise identical conditions, as follows.
\begin{gather}
	\lambda_\text{n} = \left(\frac{\lambda - \lambda_{\text{max}}}{\lambda_{\text{max}} - \lambda_{\text{min}}}  \right);
	\qquad\text{and}\qquad
	\lambda_\text{N} = \frac{\lambda}{\lambda_{\text{ref}}}; 
	%= \frac{\lambda}{\lambda(d_\text{c}=1)}
	\qquad\text{here}\qquad \lambda = (U, n^{\ast}, E_{\text{x}}, P, Y)
	\label{eq:11a}
\end{gather}
where the subscripts `max', `min' and `ref' indicate for the maximum, minimum and reference (i.e., $d_\text{c}=1$)  values of the quantity ($\lambda$) for similar flow conditions ($K, S, d_\text{c}$).
%
%--------------------------------------
\section{Numerical approach}
\label{sec:sanp}
\noindent 
In this study, the coupled mathematical model equations (\eqns\ref{eq:2} to \ref{eq:9})  have been solved using the finite element method (FEM) based COMSOL multiphysics software to obtain the electroviscous flow (EVF) fields in the charged contraction-expansion slit microfluidic device. While the detailed numerical approach is presented in recent studies  \citep{dhakar2023cfd,dhakar2022electroviscous}, only the salient features are included here to avoid duplication. The \textit{Electrostatics} (es), \textit{transport of dilute species} (tds), and \textit{laminar flow} (spf) modules of COMSOL are used to implement the multiphysics governed by the Poisson's (\eqn\ref{eq:2}), Nernst-Planck (\eqn\ref{eq:3}), and Navier-Stokes (\eqns\ref{eq:4} - \ref{eq:5}) equations, respectively. The relevant boundary conditions (\eqns\ref{eq:6} to \ref{eq:9}) are also implemented through these multiphysics modules (es, tds, spf). Further, the integral quantities in the current continuity condition (\eqn\ref{eq:7}) are obtained by using  \textit{intop} function in the model coupling through the `global function definition'. The mathematical model equations (\eqns\ref{eq:2} to \ref{eq:9}) and geometrical problem domain are discretized using the finite element method (FEM). The resulting set of coupled algebraic equations is solved iteratively using the PARDISO (PARallel DIrect SOlver), Newton's non-linear solvers, and fully coupled method to obtain the steady-state solution for the total electrical potential ($U$), induced electric field strength ($E_\text{x}$), ionic concentration ($n_{\pm}$), velocity ($\myvec{V}$), and pressure ($P$) fields.
%
%\\\noindent
The dimensionless geometrical characteristics (\fig\ref{fig:1}) are systematically optimized and obtained \citep{davidson2007electroviscous,bharti2008steady,dhakar2022electroviscous} as follows:  $L_\text{u} = L_\text{c} = L_\text{d} = 5$, $L = 15$, and $0.25\le W_\text{c}\le 1$. Further, the sufficiently refined mesh, free from end effects, providing most accurate results is obtained \citep{dhakar2022electroviscous} as follows: $\Delta = 100$, correspondingly  produces ($N_e$, DoF) as (333600, 3018814), (358400,	3238126), (383200,	3462776) and (408000, 3687434) for $d_\text{c}$ of 0.25, 0.50, 0.75, and 1, respectively, where $\Delta$ is number of mesh points uniformly distributed per unit length of the boundaries, $N_e$ is the total number of mesh elements in the computational domain  (\fig\ref{fig:1}), and DoF is the degrees of freedom. 
%--------------------------------------
%
%---------------------------------------------------------------
\section{Results and discussion}
%---------------------------------------------------------------
%
\noindent 
In this section, the detailed electroviscous flow characteristics have been obtained and presented for the pressure-drive flow of electrolyte liquid in the symmetrically charged non-uniform microfluidic device by systematic variation of the following flow conditions such as surface charge density ($4\le S\le 16$), inverse Debye length ($2\le K\le 20$), and contraction ratio ($0.25\le d_\text{c}\le 1$) for fixed values of  Schmidt and Reynolds numbers ($\mathit{Sc}=10^3$; $Re=10^{-2}$). 
In general, the functional dependence of flow conditions ($K$, $S$, $d_\text{c}$) on the total electrical potential ($U$), excess ionic charge ($n^\ast$), axially induced electric field strength ($E_\text{x}$), pressure ($P$) fields, and electroviscous correction factor ($Y$) are presented and discussed. 
The considered range of conditions ($K$, $S$) are justified \citep{davidson2007electroviscous,bharti2008steady,bharti2009electroviscous,dhakar2022electroviscous,dhakar2023cfd} for their practical applicability, and discussion is not repeated here.  
The numerical approach used in this work has been validated \citep{dhakar2022electroviscous} with the limiting literature \citep{davidson2007electroviscous} for the fixed contraction ratio ($d_\text{c}=0.25$)  over the ranges of other conditions ($K, S$). Based on our experience \citep{bharti2008steady,bharti2009electroviscous,dhakar2022electroviscous,dhakar2023cfd},  the results presented hereafter are reliable and accurate within $\pm 1-2\%$.
\subsection{Total electrical potential ($U$) distribution}
%%%%%%%%%%
\label{sec:potential}
\noindent 
The distribution of total electrical potential ($U$) along the centreline ($0\le x\le L$, 0)  of the uniformly charged microfluidic device as a function of governing parameter ($K$, $S$, $d_\text{c}$) is qualitatively similar (\fig\ref{fig:2}) to that reported in the literature \citep{davidson2007electroviscous,bharti2008steady,dhakar2022electroviscous,dhakar2023cfd} for no-slip flow through fixed contraction-expansion ($d_\text{c}=0.25$) microfluidic device. For instance,  the potential decreases along the length of the device (i.e., $U \propto x^{-1}$), inversely proportional to the Debye parameter ($U \propto K^{-1}$), and proportional to the surface charge density ($U \propto S$), irrespective of the contraction size ($d_\text{c}$). Further, the potential drop is highest in the contraction region compared to other regions.

%Normalized total electrical potential is defined as $U_\text{n}=\frac{U-U_\text{max}}{U_\text{max}-U_\text{min}}$, here $U_\text{max}$ and $U_\text{min}$ are maximum and minimum value of potential for each $K$, respectively. 
%
\begin{figure}[h]
	\centering\includegraphics[width=1\linewidth]{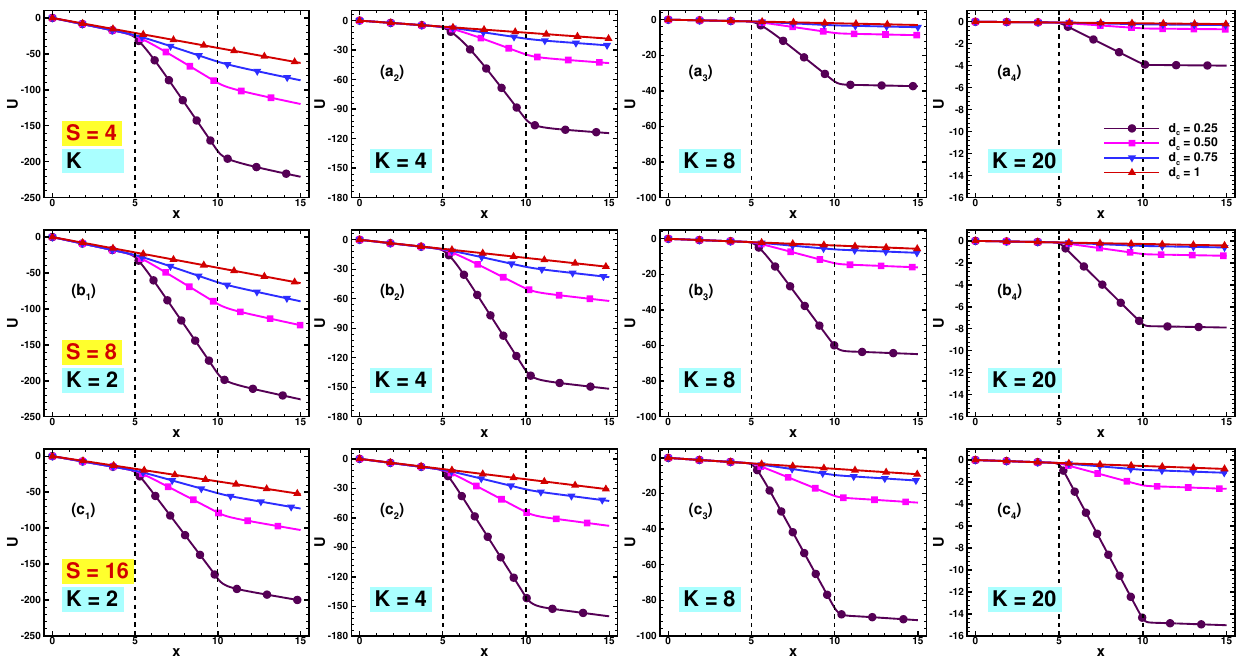}
	\caption{Total electrical potential ($U$) on the centreline ($0\le x\le L$, 0) of charged microfluidic device as a function of dimensionless governing parameters ($K$, $S$, $d_\text{c}$).}
	\label{fig:2}
\end{figure}
\begin{figure}[t]
	\centering\includegraphics[width=1\linewidth]{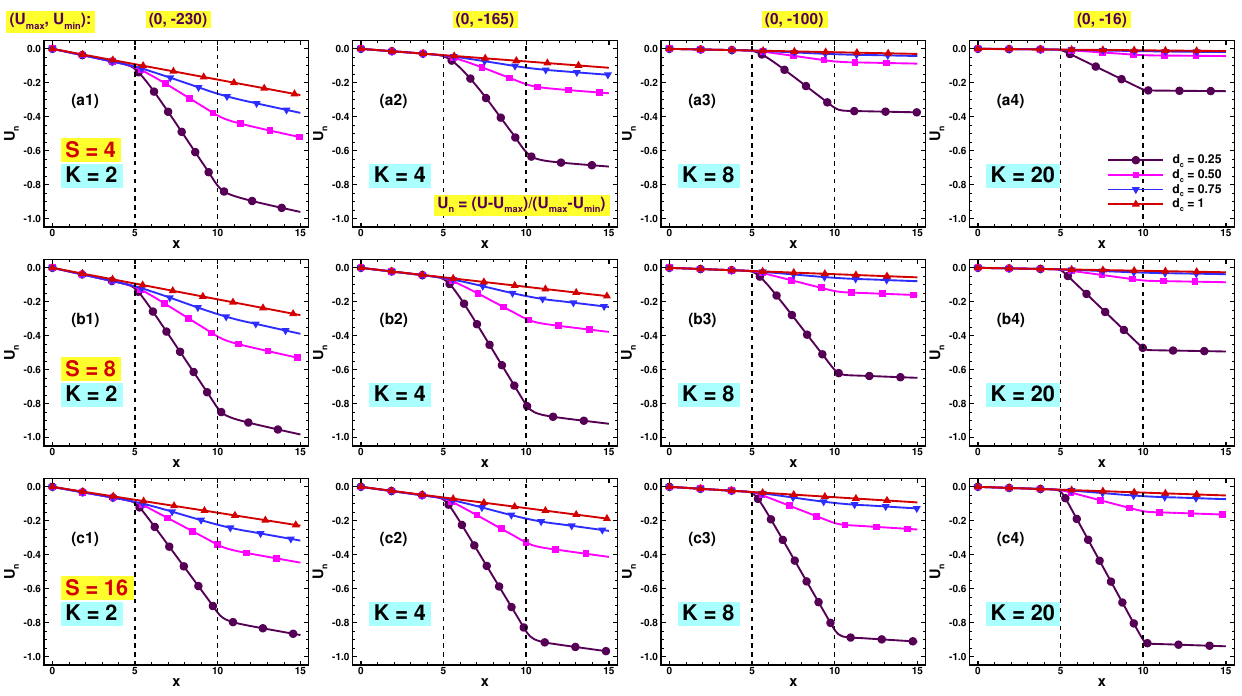}
	\caption{Centreline profiles of normalized total electrical potential ($U_\text{n}$) for $2\le K\le 20$, $4\le S\le 16$, and $0.25\le d_\text{c}\le 1$.}
	\label{fig:2c}
\end{figure}

 
\fig\ref{fig:2c} depicts centreline profiles of normalized total electrical potential ($U_\text{n}$) in the charged microchannel for $2\le K\le 20$, $4\le S\le 16$, and $0.25\le d_\text{c}\le 1$. Centreline profiles of $U_\text{n}$ have shown similar qualitative behavior as $U$ with the literature \citep{davidson2007electroviscous,dhakar2022electroviscous,dhakar2023cfd}. Along the length of positively charged device, $U_\text{n}$ decreases due to advection of excess charge (negative ions) occurs as expected in the PDF direction. Normalized potential gradient is maximum in the contraction as compared to other sections (\fig\ref{fig:2c}). Normalized potential decreases with decreasing $K$ or EDL thickening (\fig\ref{fig:2c}). The variation of $U_\text{n}$ with $K$ is maximum at higher $d_\text{c}$ and lower $S$. For instance, $|U_\text{n}|$ reduces maximally by 95.14\% (0.2721 to 0.0132) when $K$ varies from 2 to 20 at $S=4$ and $d_\text{c}=1$ (refer \fig\ref{fig:2c}a). Normalized potential decreases with increasing $S$ (\fig\ref{fig:2c}a and b) but at higher $S$, enhances with increasing $S$ (\fig\ref{fig:2c}c). For instance, $|U_\text{n}|$ increases maximally by 285.24\% (0.0132 to 0.0510) when $S$ varies from 4 to 16 at $K=20$ and $d_\text{c}=1$ (refer \fig\ref{fig:2c}). Further, $|U_\text{n}|$ increases with decreasing $d_\text{c}$ (\fig\ref{fig:2c}). The relative impact of $d_\text{c}$ on $U_\text{n}$ is maximum at higher $K$ and lower $S$ because EDLs do not occupy a greater fraction of the microfluidic device. For instance, $|U_\text{n}|$ enhances maximally by 1785.58\% (0.0132 to 0.2496) when $d_\text{c}$ varies from 1 (uniform device) to 0.25 (contraction-expansion device) at weak-EVF condition ($K=20$, $S=4$) (refer \fig\ref{fig:2c}a4).
\\\noindent Subsequently, \tab\ref{tab:1} comprises total potential drop ($|\Delta U|$) on the centreline ($0\le x\le L$, $0$) of microfluidic device for wide ranges of dimensionless parameters ($2\le K\le 20$, $4\le S\le 16$, $0.25\le d_\text{c}\le 1$). Maximum value of potential drop ($|\Delta U|_\text{max}$) is underline for $0.25\le d_\text{c}\le 1$ at each combination of $S$ and $K$. The variation in the value of $|\Delta U|$ with $K$ and $S$ for a fixed $d_\text{c}=0.25$ is similar as the literature \citep{davidson2007electroviscous,dhakar2022electroviscous,dhakar2023cfd}. Potential drop ($|\Delta U|$) decreases with increasing $K$ (or EDL thinning); the decrement of $|\Delta U|$ with $K$ is lesser for higher $S$ and lower $d_\text{c}$ (\tab\ref{tab:1}). For instance, $|\Delta U|$ reduces by (98.19\%, 99.43\%, 99.65\%, 99.66\%) and (92.51\%, 97.45\%, 98.40\%, 98.45\%) for ($d_\text{c}=0.25$, 0.50, 0.75, 1), respectively at $S=4$ and 16 when $K$ varies from 2 to 20 (refer \tab\ref{tab:1}). The $|\Delta U|$ increases with increasing $S$ followed by reverse trends at higher $S$; it attributes due to strong charge attractive forces near the surface retards the convective flow of ions in device at higher $S$. The relative impact of $S$ on $|\Delta U|$ is maximum at higher $K$ and $d_\text{c}$; it is because EDLs overlap at lower $K$ and higher $S$ (\tab\ref{tab:1}). For instance, $|\Delta U|$ varies with increasing $S$ from 4 to 16 by (-9.14\%, -14.19\%, -16.07\%, -15.90\%) and (276.44\%, 282.83\%, 284.63\%, 285.24\%) for ($d_\text{c}=0.25$, 0.50, 0.75, 1), respectively at $K=2$ and 20 (refer \tab\ref{tab:1}). Further, $|\Delta U|$ enhances with decreasing $d_\text{c}$ because reduction in the cross-section flow area increases excess charge clustering and velocity in the microfluidic device , thus, increment in the streaming current decreases streaming potential (\tab\ref{tab:1}). For instance, $|\Delta U|$ enhances for ($S=4$, 8, 16) by (38.66\%, 39.50\%, 38.37\%) and (43.20\%, 43.15\%, 42.98\%) at $K=2$ and 20, respectively with decreasing contraction $d_\text{c}$ from 1 to 0.75; corresponding increment in the values of $|\Delta U|$ for decreasing $d_\text{c}$ from 1 to 1.50 are recorded as (91.31\%, 91.42\%, 95.19\%) and (223.49\%, 223.06\%, 221.47\%) at $K=2$ and 20. Similarly, $|\Delta U|$ enhances at $K=2$ and 20 by (253.07\%, 252.36\%, 281.45\%) and (1785.58\%, 1776.40\%, 1742.51\%) with overall decreasing contraction $d_\text{c}$ from 1 to 0.25 ($0.25\le d_\text{c}\le 1$) (refer \tab\ref{tab:1}).
%Therefore, it is found that variation in the value of potential drop is maximum at higher $K$ and lower $S$ at fixed $d_\text{c}$ (as shown in \tab\ref{tab:1}). It is because EDLs do not occupy a greater fraction of microchannel at higher $K$ and lower $S$, therefore, potential drop changes maximum at higher $K$ and lower $S$. Further, the change in the value of potential drop is greater when $d_\text{c}$ varies from $1$ to $0.25$ as compared with from $1$ to $0.75$ at fixed $K$ and $S$ (as shown in \tab\ref{tab:1}). It is because reduction in the cross-section area enhances the excess charge clustering and velocity in that section, therefore, increment in the streaming current decrease the total potential.
%
%
\begin{table}[!tb]
	\centering
	\caption{Total electrical potential drop ($|\Delta U|$), minimum value of excess charge ($n^\ast_{\text{min}}$), maximum value of induced electric field strength ($E_\text{x,max}$), and pressure drop ($10^{-5}|\Delta P|$) on the centreline of device for $2\le K\le 20$, $4\le S\le 16$, and $0.25\le d_\text{c}\le 1$.}\label{tab:1}
	\scalebox{0.8}
	{
		\begin{tabular}{|r|r|r|r|r|r|r|r|r|r|}%r|r|r|r|r|r|r|r|}
		\hline
		$S$	&	$K$	&	$d_\text{c}=0.25$	&	$0.50$	& $0.75$ &	$1$ &	$d_\text{c}=0.25$	&	$0.50$	& $0.75$ &	$1$ 
		\\\cline{3-10}%\cline{3-18}
		&	&	\multicolumn{4}{c|}{$\Delta U$}	&	\multicolumn{4}{c|}{$n^\ast_{\text{min}}$} 
		\\\hline
		0   & 	$\infty$    & 	0	    & 0	        & 0          & 	0 & 	0	    & 0	        & 0          & 	0 
		\\\hline
		4	&	2	&	\underline{220.9700}	&	119.7300	&	86.7780	&	62.5850	&	\underline{-6.6597}	&	-2.7279	&	-1.4805	&	-0.8159	
		\\
		&	4	&	\underline{114.4300}	&	43.1980	&	25.3040	&	18.4210	&	\underline{-1.5275}	&	-0.4923	&	-0.1809	&	-0.0662	
		\\
		&	6	&	\underline{63.6860}	&	17.8980	&	9.2466	&	6.6066	&	\underline{-0.5902}	&	-0.1267	&	-0.0283	&	-0.0063	
		\\
		&	8	&	\underline{37.4060}	&	8.7252	&	4.2351	&	2.9985	&	\underline{-0.2669}	&	-0.0357	&	-0.0048	&	-0.0007	
		\\
		&	20	&	\underline{3.9929}	&	0.6850	&	0.3033	&	0.2118	&	\underline{-5.3110$\times10^{-3}$}	&	-3.5410$\times10^{-5}$	&	-2.3732$\times10^{-7}$	&	-1.6119$\times10^{-9}$	
		\\\hline
		8	&	2	&	\underline{225.7700}	&	122.6500	&	89.3850	&	64.0740	&	\underline{-11.6680}	&	-4.3860	&	-2.2448	&	-1.1391	
		\\
		&	4	&	\underline{151.5500}	&	62.2770	&	38.0280	&	27.7130	&	\underline{-2.7181}	&	-0.8280	&	-0.3015	&	-0.1072	
		\\
		&	6	&	\underline{99.3690}	&	30.5710	&	16.2210	&	11.6700	&	\underline{-1.0665}	&	-0.2271	&	-0.0511	&	-0.0113	
		\\
		&	8	&	\underline{64.8040}	&	16.0670	&	7.9006	&	\underline{-5.6197}	&	-0.4960	&	-0.0669	&	-0.0091	&	-0.0012	
		\\
		&	20	&	\underline{7.8852}	&	1.3576	&	0.6016	&	0.4202	&	\underline{-1.0522$\times10^{-2}$}	&	-7.0216$\times10^{-5}$	&	-4.6787$\times10^{-7}$	&	-3.1843$\times10^{-9}$	
		\\\hline
		16	&	2	&	\underline{200.7800}	&	102.7400	&	72.8350	&	52.6360	&	\underline{-18.3170}	&	-6.1084	&	-2.9253	 &	-1.3549	
		\\
		&	4	&	\underline{160.1100}	&	68.1120	&	42.8220	&	31.1170	&	\underline{-4.2938}	&	-1.1887	&	-0.4199	&	-0.1400	
		\\
		&	6	&	\underline{122.1500}	&	41.4390	&	22.9800	&	16.7180	&	\underline{-1.7149}	&	-0.3485	&	-0.0784	&	-0.0170	
		\\
		&	8	&	\underline{91.1770}	&	25.1300	&	12.7710	&	9.1889	&	\underline{-0.8248}	&	-0.1109	&	-0.0151	&	-0.0020	
		\\
		&	20	&	\underline{15.0310}	&	2.6225	&	1.1664	&	0.8158	&	\underline{-2.0306$\times10^{-2}$}	&	-1.3595$\times10^{-4}$	&	-9.0796$\times10^{-7}$	&	-6.1265$\times10^{-9}$	
		\\\hline
		&	&	\multicolumn{4}{c|}{$E_\text{x,max}$} &   \multicolumn{4}{c|}{$10^{-5}|\Delta P|$} 
		\\\hline%\cline{3-18}
		0   & 	$\infty$   & 	0	    & 0	        & 0          & 0     & 1.0616      & 0.1624	        & 0.0678          & 	0.0450	 
		\\\hline
		4	&	2	&	\underline{33.6790}	&	13.9680	&	7.6907	&	4.1720	&	\underline{1.1673}	&	0.1896	&	0.0818	&	0.0533	
		\\
		&	4	&	\underline{22.1080}	&	6.2858	&	2.5460	&	1.2281	&	\underline{1.1189}	&	0.1715	&	0.0710	&	0.0468	
		\\
		&	6	&	\underline{12.9000}	&	2.7561	&	0.9681	&	0.4406	&	\underline{1.0952}	&	0.1658	&	0.0687	&	0.0455	
		\\
		&	8	&	\underline{7.4680}	&	1.3584	&	0.4502	&	0.2000	&	\underline{1.0811}	&	0.1638	&	0.0681	&	0.0451	
		\\
		&	20	&	\underline{0.7630}	&	0.1074	&	0.0332	&	0.0142	&	\underline{1.0629}	&	0.1624	&	0.0678	&	0.0450	
		\\\hline
		8	&	2	&	\underline{33.7070}	&	14.1030	&	7.8680	&	4.2707	&	\underline{1.2767}	&	0.2128	&	0.0926	&	0.0594	
		\\
		&	4	&	\underline{26.4830}	&	8.5533	&	3.7429	&	1.8475	&	\underline{1.2136}	&	0.1869	&	0.0767	&	0.0501	
		\\
		&	6	&	\underline{18.8840}	&	4.5930	&	1.6844	&	0.7782	&	\underline{1.1640}	&	0.1735	&	0.0709	&	0.0467	
		\\
		&	8	&	\underline{12.5870}	&	2.4787	&	0.8372	&	0.3749	&	\underline{1.1274}	&	0.1676	&	0.0690	&	0.0456	
		\\
		&	20	&	\underline{1.5062}	&	0.2128	&	0.0658	&	0.0281	&	\underline{1.0669}	&	0.1626	&	0.0678	&	0.0450	
		\\\hline
		16	&	2	&	\underline{30.5860}	&	12.0150	&	6.4612	&	3.5069	&	\underline{1.4180}	&	0.2330	&	0.0996	&	0.0633	
		\\
		&	4	&	\underline{26.6960}	&	8.9506	&	4.1103	&	2.0741	&	\underline{1.3579}	&	0.2082	&	0.0845	&	0.0544	
		\\
		&	6	&	\underline{21.8890}	&	5.9714	&	2.3398	&	1.1148	&	\underline{1.2929}	&	0.1889	&	0.0755	&	0.0492	
		\\
		&	8	&	\underline{16.9910}	&	3.7871	&	1.3386	&	0.6131	&	\underline{1.2329}	&	0.1772	&	0.0715	&	0.0469	
		\\
		&	20	&	\underline{2.8667}	&	0.4105	&	0.1276	&	0.0546	&	\underline{1.0815}	&	0.1632	&	0.0679	&	0.0450	
		\\\hline
	\end{tabular}
}
\end{table}
\begin{figure}[t]
\centering\includegraphics[width=1\linewidth]{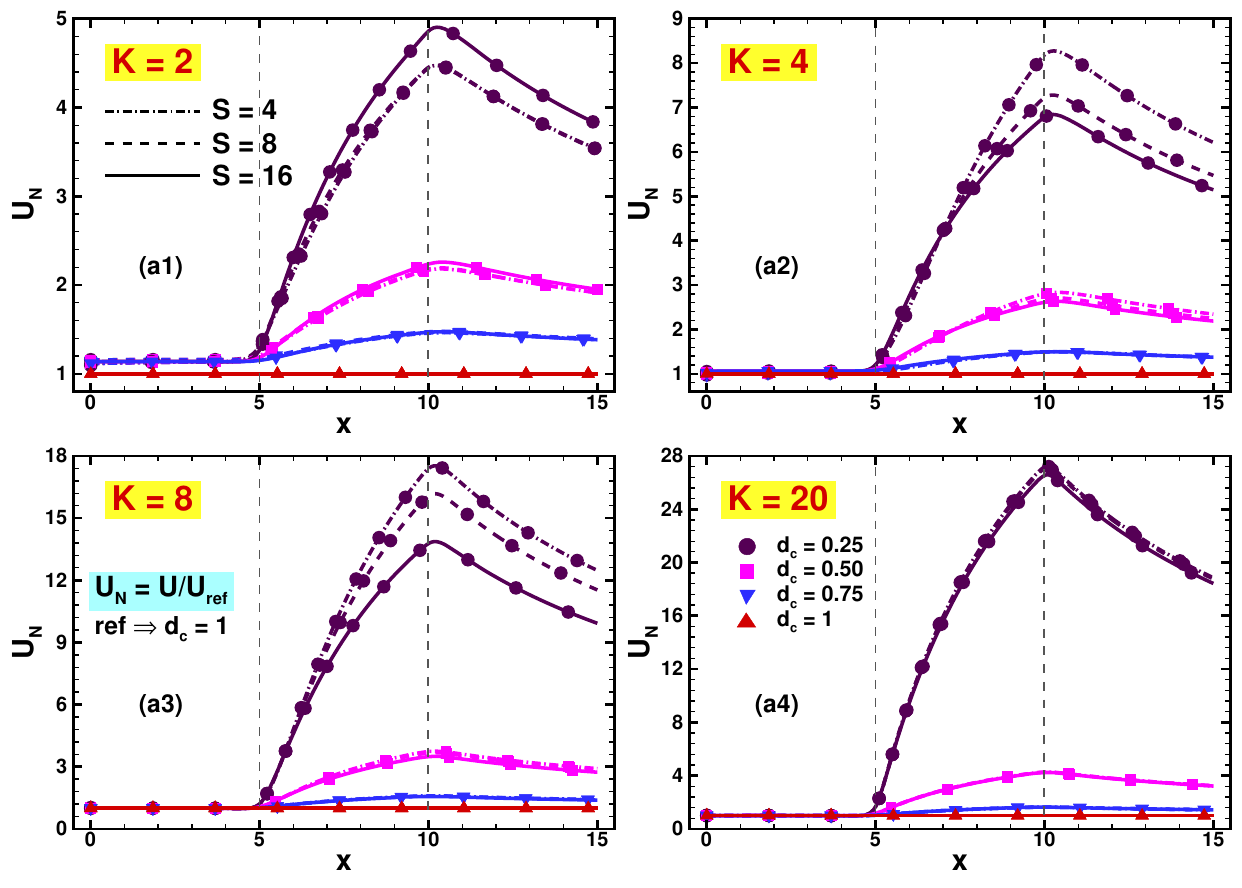}
\caption{Centreline profiles of normalized total electrical potential ($U_\text{N}$) for $2\le K\le 20$, $4\le S\le 16$, and $0.25\le d_\text{c}\le 1$.}
\label{fig:2d}
\end{figure} 
\\\noindent Further, the relative impact of contraction ($d_\text{c}$) is described by normalizing the flow fields for non-uniform ($d_\text{c}\neq1$) geometries by that at reference case of uniform (`ref' or $d_\text{c}=1$) geometry for dimensionless parameters ($S$, $K$) as given below.
\begin{gather}
\theta_\text{N} = \frac{\theta}{\theta_{\text{ref}}} = 
\left.\frac{\theta(d_\text{c})}{\theta(d_\text{c}=1)}\right|_{S, K}
\qquad\text{where}\qquad \theta = (U, n^{\ast}, E_{\text{x}}, P, Y)
\label{eq:11a}
\end{gather}%%%
\noindent For detailed analysis of total potential variation on centreline ($0\le x\le L$, 0) of device, it is normalized by reference potential (at $d_\text{c}=1$). \fig\ref{fig:2d} shows centreline profiles of normalized total electrical potential ($U_\text{N}$, \eqn\ref{eq:11a}) in the charged microchannel for $2\le K\le 20$, $4\le S\le 16$, and $0.25\le d_\text{c}\le 1$. At $d_\text{c}<1$, normalized potential ($U_\text{N}$) is constant in the upstream section and increases drastically in contraction section, followed by continuously decrease in downstream region. However, $U_\text{N}$ is unity along the length at $d_\text{c}=1$ due to normalized by this condition (\fig\ref{fig:2d}). The increment in the value of $U_\text{N}$ is noted with increasing $K$ because maximum change in the value of $U$ with $d_\text{c}$ is obtained at higher $K$, irrespective of $S$ (refer \tab\ref{tab:1}). The relative effect of $K$ on $U_\text{N}$ is maximum at lower $d_\text{c}$ and $S$. For instance, $U_\text{N}$ maximally enhances by 434.05\% (at $S=4$, $d_\text{c}=0.25$) when $K$ varies from 2 to 20 (refer \fig\ref{fig:2d}). Normalized potential increases with decreasing $S$ (\fig\ref{fig:2d}a2 to a4); however, it increases with increasing $S$ at lower $K=2$ (thick EDL) (\fig\ref{fig:2d}a1). It is because EDLs overlap at lower $K$ and higher $S$. The variation in the value of $U_\text{N}$ with $S$ is maximum at $K=8$. For instance, $U_\text{N}$ reduces by 7.56\% (12.4749 to 11.5316) and 20.46\% (12.4749 to 9.9225) for increasing $S$ (from 4 to 8) and (from 4 to 16), respectively at $d_\text{c}=0.25$ and $K=8$ in the end location ($x=10$) of contraction section (refer \fig\ref{fig:2d}a3). Further, $U_\text{N}$ enhances with decreasing $d_\text{c}$ (\fig\ref{fig:2d}) due to increase $|\Delta U|$ with reducing cross-section flow area of middle section (refer \tab\ref{tab:1}). It is because dense clustering of excess charge and velocity enhancement in that section, therefore, decrease streaming potential and increase normalized total potential. Further, the variation in the value of $U_\text{N}$ with $d_\text{c}$ is maximum at higher $K$ and lower $S$ because EDLs are thinner and do not cover greater faction of device at this condition (\fig\ref{fig:2d}). 
%%
%For instance, maximum enhancement in $U_\text{N}$ is recorded as 1785.58\% (1 to 18.8558) for overall reduction in the contraction $d_\text{c}$ from 1 to 0.25 ($0.25\le d_\text{c}\le 1$) at $K=20$, $S=4$ (refer \fig\ref{fig:2d}a4).
%%
%
\\\noindent Further, total potential ($U$) distribution relates with excess ionic charge ($n^\ast$) by Poisson's equation (\eqn\ref{eq:2}). Thus, preceding section present the excess ionic charge ($n^\ast$) as a function of governing parameters ($S$, $K$, $d_\text{c}$). 
%
%
%
%%%%%%%%%%%%%%%
\subsection{Excess charge ($n^\ast$) distribution}
\begin{figure}[t]
\centering\includegraphics[width=1\linewidth]{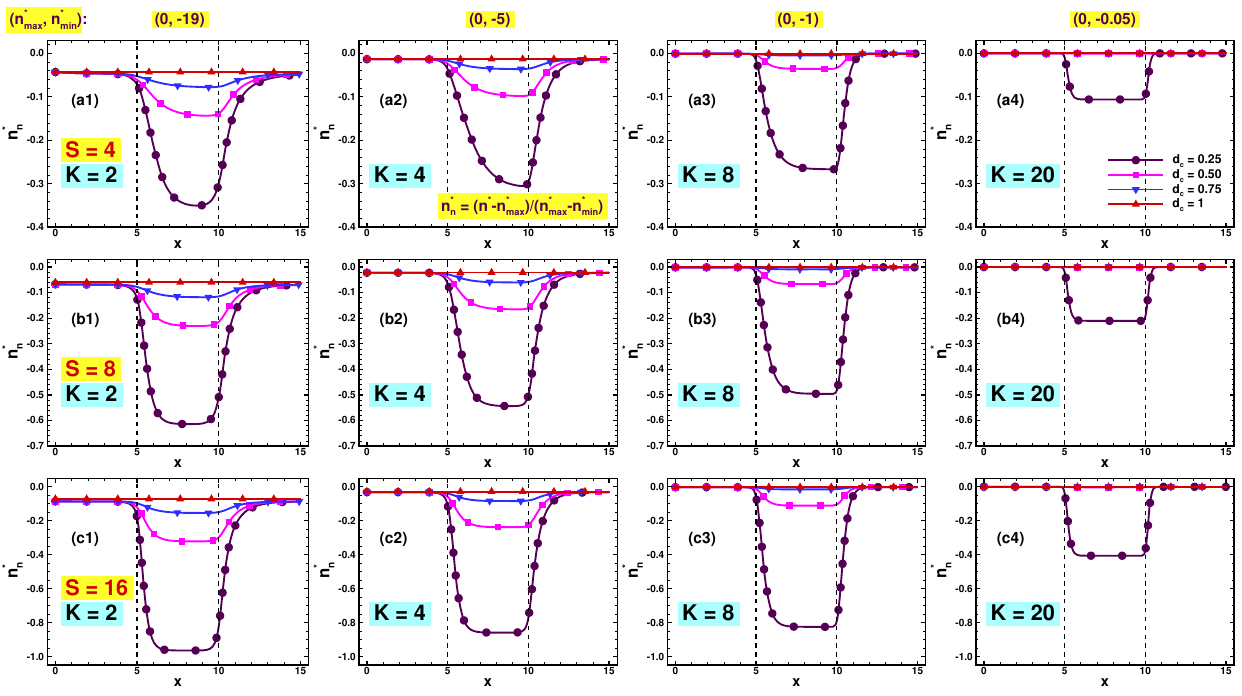}
\caption{Centreline profiles of normalized excess charge ($n^\ast_{\text{n}}$) for $2\le K\le 20$, $4\le S\le 16$, and $0.25\le d_\text{c}\le 1$.}
\label{fig:3b}
\end{figure}

%%%%%%%%%%%%%%%
\label{sec:charge}
%%%%%%%%%%%%%%%
%
\noindent Normalized excess charge is defined as $n^\ast_{\text{n}}=\frac{n^\ast-n^\ast_{\text{max}}}{n^\ast_{\text{max}}-n^\ast_{\text{min}}}$, here $n^\ast_\text{min}$ and $n^\ast_\text{max}$ represent the minimum and maximum values of $n^\ast$ for each inverse Debye length ($K$). \fig\ref{fig:3b} shows centreline profiles of normalized excess charge $n^\ast_{\text{n}}$ in the charged microchannel for $2\le K\le 20$, $4\le S\le 16$, and $0.25\le d_\text{c}\le 1$. Qualitative behavior of normalized centreline excess charge profiles are same as $n^\ast $ with the literature \citep{davidson2007electroviscous,dhakar2022electroviscous,dhakar2023cfd}. Normalized excess charge ($n^\ast_{\text{n}}$) is equal and most prominent at horizontal centreline of channel ($0\le x\le L$, 0) of upstream and downstream sections in the device (\fig\ref{fig:3b}). The $n^\ast_\text{n}$ is minimum in the contraction than other sections of channel (\fig\ref{fig:3b}). Minimum value of normalized excess charge ($n^\ast_{\text{n,min}}$) increases with increasing $K$ or EDL thinning; $n^\ast_\text{n,min}$ values are close to zero at higher $K=20$ (\fig\ref{fig:3b}). For instance, $n^\ast_{\text{n,min}}$ enhances maximally by 100\% (-0.0429 to -3.2238$\times10^{-8}$), 100\% (-0.0600 to -6.3686$\times10^{-8}$), and 100\% (-0.0713 to -1.2253$\times10^{-7}$) for $S=4$, 8, and 16, respectively when $K$ varies from 2 to 20 at $d_\text{c}=1$ (refer \fig\ref{fig:3b}). The $n^\ast_{\text{n,min}}$ decreases with increasing $S$ (\fig\ref{fig:3b}). For instance, $n^\ast_{\text{n,min}}$ decreases maximally by 283.93\% (-0.0007 to -0.0027) when $S$ varies from 4 to 16 at $K=20$ and $d_\text{c}=0.5$ (refer \fig\ref{fig:3b}). Further, $n^\ast_{\text{n,min}}$ decreases with decreasing $d_\text{c}$ (\fig\ref{fig:3b}); the variation in the value of $n^\ast_{\text{n,min}}$ with $d_\text{c}$ is greater at higher $K$ and $S$ (\fig\ref{fig:3b}). It is because EDLs are very thin in the device at higher $K$ and available excess ions ($n^\ast$) in he EDL are more at higher $S$. For instance, $n^\ast_{\text{n,min}}$ reduces maximally by 3.31$\times10^8\%$ (-1.2253$\times10^{-7}$ to -0.4061) when $d_\text{c}$ varies from 1 to 0.25 at $K=20$ and $S=16$ (refer \fig\ref{fig:3b}c4). 
\\\noindent Subsequently, excess charge ($n^\ast$) is minimum in the contraction than other sections of device. \tab\ref{tab:1} summarizes the minimum excess charge ($n^\ast_{\text{min}}$) on the centreline ($0\le x\le L$, 0) of microfluidic device for $2\le K\le 20$, $4\le S\le 16$, and $0.25\le d_\text{c}\le 1$. The variation of $n^\ast_{\text{min}}$ with $K$ and $S$ for a fixed $d_\text{c}=0.25$ is similar as the literature \citep{davidson2007electroviscous,dhakar2022electroviscous, dhakar2023cfd}. The $n^\ast_\text{min}$ increases with increasing $K$ (or EDL thinning); at higher inverse Debye length ($K>20$), $n^\ast_\text{min}$ tends to be zero (\tab\ref{tab:1}). For instance, $n^\ast_{\text{min}}$ reduces for ($d_\text{c}=0.25$, 0.50, 0.75, 1) by (99.92\%, 100\%, 100\%, 100\%) and (99.89\%, 100\%, 100\%, 100\%), respectively at $S=4$ and 16 when $K$ varies from 2 to 20 (refer \tab\ref{tab:1}). The $n^\ast_\text{min}$ decreases with increasing $S$ because enhancement in the electrostatic forces near the device walls, increases clustering of excess ions in the channel (\tab\ref{tab:1}). The relative effect of $S$ on $n^\ast_\text{min}$ is maximum at higher $K$. For instance, increment in the value of $n^\ast_{\text{min}}$ are recorded with $S$ varies from 4 to 16 as (175.04\%, 123.92\%, 97.59\%, 66.07\%) and (282.34\%, 283.93\%, 282.59, 280.08\%) for ($d_\text{c}=0.25$, 0.50, 0.75, 1), respectively at $K=2$ and 20. Further, $|n^\ast_{\text{min}}|$ enhances with decreasing $d_\text{c}$ because reduction in the cross-flow area of middle section results dense clustering of excess (negative) ions in the microfluidic device (\tab\ref{tab:1}). The variation of $n^\ast_{\text{min}}$ with $d_\text{c}$ is maximum at higher $K$ and $S$. For instance, $n^\ast_{\text{min}}$ enhances for ($S=4$, 8, 16) by (81.47\%, 97.07\%, 115.91\%) and (1.46$\times10^4\%$, 1.46$\times10^4\%$, 1.47$\times10^4\%$), respectively at $K=2$ and 20 with decreasing $d_\text{c}$ from 1 to 0.75; corresponding increment in the values of $n^\ast_{\text{min}}$ with decreasing $d_\text{c}$ from 1 to 0.50 are recorded as (234.36\%, 285.04\%, 350.84\%) and (2.20$\times10^6\%$, 2.20$\times10^6\%$, 2.22$\times10^6\%$) at $K=2$ and 20. Similarly, $n^\ast_{\text{min}}$ enhances at $K=2$ and 20 by (716.29\%, 924.32\%, 1251.91\%) and (3.29$\times10^8\%$, 3.30$\times10^8\%$, 3.31$\times10^8\%$) with overall decreasing contraction $d_\text{c}$ from 1 to 0.25 ($0.25\le d_\text{c}\le 1$) (refer \tab\ref{tab:1}).
\begin{figure}[t]
\centering\includegraphics[width=1\linewidth]{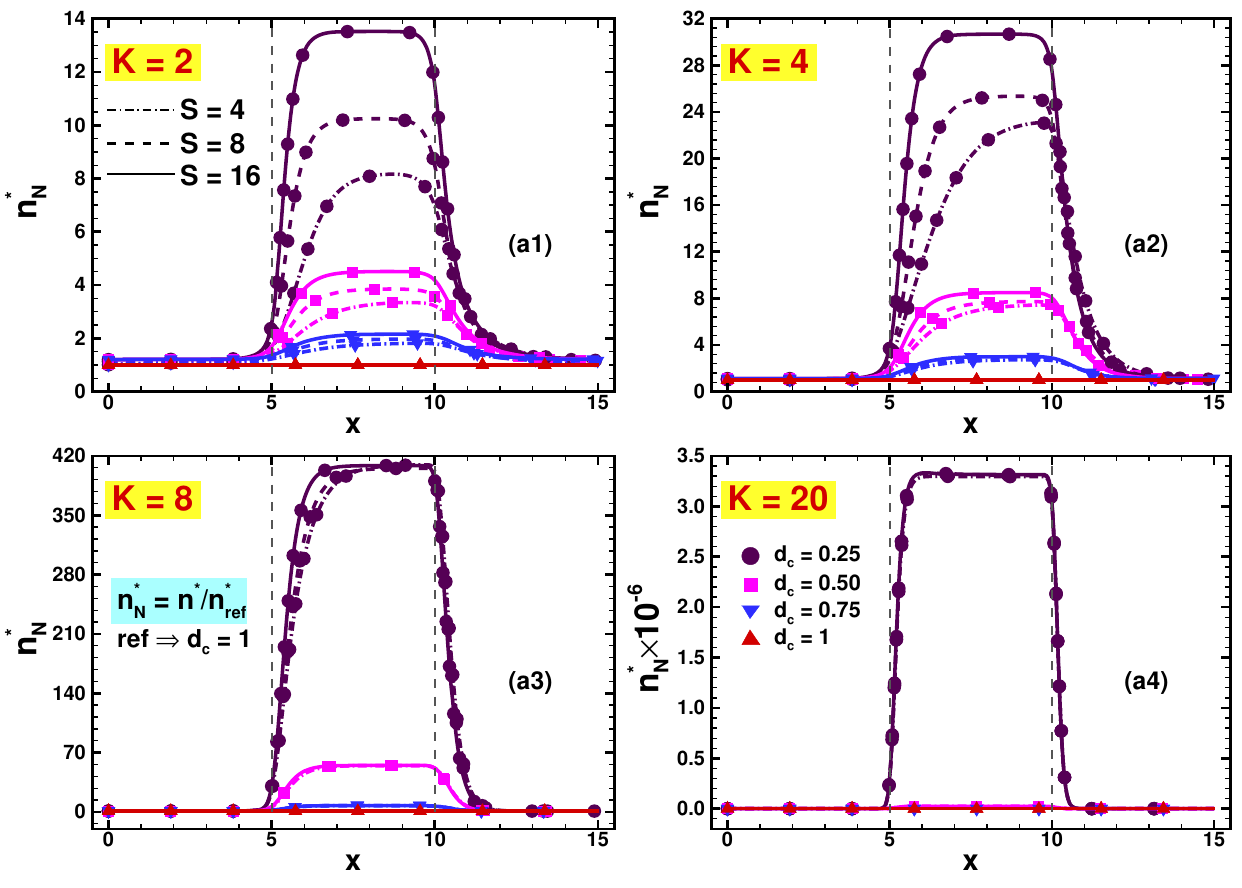}
\caption{Centreline profiles of normalized excess charge ($n^\ast_{\text{N}}$) for $2\le K\le 20$, $4\le S\le 16$, and $0.25\le d_\text{c}\le 1$.}
\label{fig:3c}
\end{figure} 
\\\noindent Further, the detailed variation of the excess charge ($n^\ast$) on the centreline ($0\le x\le L$, 0) of device is explained by normalizing the excess charge with reference case (at $d_\text{c}=1$). \fig\ref{fig:3c} shows centreline profiles of normalized excess charge ($n^\ast_{\text{N}}$, \eqn\ref{eq:11a}) in the charged microchannel for $2\le K\le 20$, $4\le S\le 16$, and $0.25\le d_\text{c}\le 1$. Normalized excess charge ($n^\ast_{\text{N}}$) is constant in upstream and downstream section over the horizontal centreline of device. Normalized excess charge ($n^\ast_{\text{N}}$) is maximum in the contraction than other sections for $d_\text{c}\neq1$ and equal to unity for $d_\text{c}=1$ in the device, respectively (\fig\ref{fig:3c}). Maximum value of normalized excess charge ($n^\ast_{\text{N,max}}$) increases with increasing $K$ (\fig\ref{fig:3c}) due to maximum variation of $n_\text{max}$ with $d_\text{c}$ is obtained at higher $K$, irrespective of $S$ (refer \tab\ref{tab:1}). For instance, $n^\ast_{\text{N,max}}$ enhances by 4.0364$\times10^7$\% (8.1629 to 3.2949$\times10^6$) when $K$ changes from 2 to 20 at $S=4$ and $d_\text{c}=0.25$ (refer \fig\ref{fig:3c}). The $n^\ast_{\text{N,max}}$ enhances with increasing $S$ (\fig\ref{fig:3c}); the relative impact of $S$ on $n^\ast_{\text{N,max}}$ is maximum at lower $K$ and $d_\text{c}$. For instance, $n^\ast_{\text{N,max}}$ enhances by 65.62\% (8.1629 to 13.5191) when $S$ varies from 4 to 16 at $d_\text{c}=0.25$ and $K=2$ in the middle location ($x=7.5$) of contraction section (refer \fig\ref{fig:3c}a1). Further, $n^\ast_{\text{N,max}}$ increases with decreasing $d_\text{c}$ (\fig\ref{fig:3c}) due to reducing cross-section area of contraction section results dens clustering of excess ions ($n^\ast$) in the contraction section (refer \tab\ref{tab:1}), therefore, increases normalized excess charge. The relative impact of $d_\text{c}$ on $n^\ast_{\text{N,max}}$ is greater at higher $K$ and $S$ because EDLs do not overlap at this condition (\fig\ref{fig:3c}). 
%%
%For instance, increment in the value of $n^\ast_{\text{N,max}}$ is recorded as 3.31$\times10^8$ (1 to 3.3145$\times10^6$) when $d_\text{c}$ varies from 1 to 0.25 at $K=20$ and $S=16$ (refer \fig\ref{fig:3c}a4).
%
%
%---------------------------------
\subsection{Induced electric field strength ($E_{\text{x}}$)}
\begin{figure}[h]
\centering\includegraphics[width=1\linewidth]{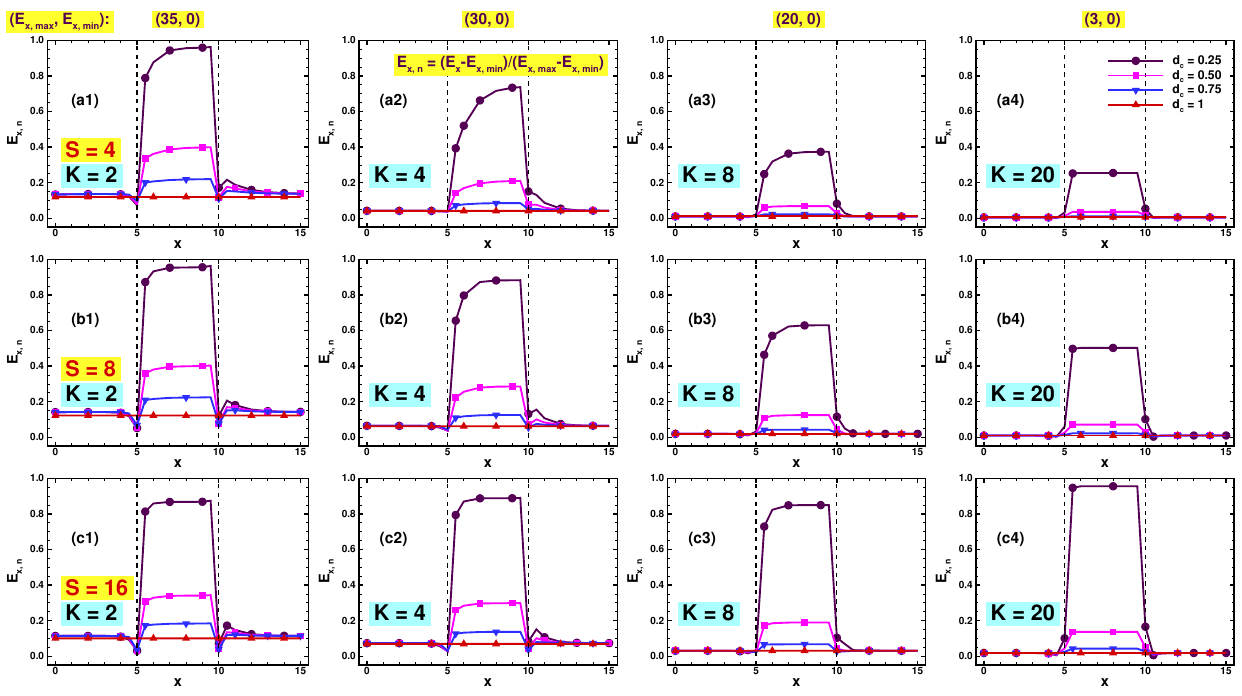}
\caption{Centreline profiles of normalized induced electric field strength ($E_\text{x,n}$) for $2\le K\le 20$, $4\le S\le 16$, and $0.25\le d_\text{c}\le 1$.}
\label{fig:4c}
\end{figure}

%---------------------------------
%%%%%%%%%%
\label{sec:electric}
%%%%%%%%%%%%%%%
%
\noindent The convective flow of ions in the charged microfluidic device develops an streaming current ($I_\text{s}$) and potential corresponding to this current is known as streaming potential. Induced electric field strength is defined as the rate of axial variation of streaming potential (i.e., $E_\text{x}=-\partial U/\partial x$) and it is calculated from \eqn(\ref{eq:7}). Normalized induced electric field strength is defined as $E_\text{x,n}=\frac{E_\text{x}-E_\text{x,min}}{E_\text{x,max}-E_\text{x,min}}$, here subscripts $\textit{max}$ and $\textit{min}$ denote the maximum and minimum values at each $K$, respectively. \fig\ref{fig:4c} depicts centreline profiles of normalized induced electric field strength ($E_\text{x,n}$) in the charged microchannel for $2\le K\le 20$, $4\le S\le 16$, and $0.25\le d_\text{c}\le 1$. Centreline profiles of $E_\text{x,n}$ depict qualitatively similar trends along the length of microfluidic device as $E_\text{x}$ with the literature \citep{dhakar2022electroviscous,dhakar2023cfd}. The $E_\text{x,n}$ is maximum in the contraction as compared the other sections (\fig\ref{fig:4c}). Maximum value of induced electric field strength ($E_\text{x,n,max}$) increases with decreasing $K$ of EDL thickening; the relative changes of $E_\text{x,n,max}$ with $K$ are maximum for lower $S$ and higher $d_\text{c}$. For instance, $E_\text{x,n,max}$ reduces maximally by 96.03\% (0.1192 to 0.0047) when $K$ varies from 2 to 20 at $S=4$ and $d_\text{c}=1$ (refer \fig\ref{fig:4c}). The $E_\text{x,n,max}$ increases with increasing $S$ followed by reverse trends at higher $S$ (\fig\ref{fig:4c}). The relative impact of $S$ is maximum on $E_\text{x,n,max}$ at higher $d_\text{c}$ and $K$. For instance, $E_\text{x,n,max}$ enhances maximally by 285.36\% (0.0047 to 0.0182) when $S$ varies from 4 to 16 at $K=20$ and $d_\text{c}=1$ (refer \fig\ref{fig:4c}). Further, $E_\text{x,n,max}$ increases with decreasing $d_\text{c}$ (\fig\ref{fig:4c}). The variation in the value of $E_\text{x,n,max}$ with $d_\text{c}$ is maximum at higher $K$ and lower $S$ because EDLs occupy the lesser fraction of device at this condition. For instance, $E_\text{x,n,max}$ enhances maximally by 5280.51\% (0.0047 to 0.2543) when $d_\text{c}$ varies from 1 to 0.25 at $K=20$ and $S=4$ (refer \fig\ref{fig:4c}a4). 
\\\noindent Further, induced electric field strength ($E_\text{x}$) is maximum in the contraction than other sections of microfluidic device. \tab\ref{tab:1} summarizes the maximum induced electric field strength ($E_\text{x,max}$) on centreline ($0\le x\le L$, 0) of microchannel for $2\le K\le 20$, $4\le S\le 16$, and $0.25\le d_\text{c}\le 1$. For a fixed $d_\text{c}=0.25$, the variation of $E_\text{x,max}$ with $S$ and $K$ is similar as the literature \citep{dhakar2022electroviscous,dhakar2023cfd}. The $E_\text{x,max}$ increases with decreasing $K$ or EDL thickening (\tab\ref{tab:1}); the relative effect of $K$ on $E_\text{x,max}$ is maximum on the lower $S$. For instance, $E_\text{x,max}$ reduces for ($d_\text{c}=0.25$, 0.50, 0.75, 1) by (97.73\%, 99.23\%, 99.57\%, 99.66\%) and (90.63\%, 96.58\%, 98.03\%, 98.44\%), respectively at $S=4$ and 16 when $K$ varies from 2 to 20 (refer \tab\ref{tab:1}). The increment in the values of $E_\text{x,max}$ are noted with increasing $S$ but at higher $S$, it attributes the reverse trends due to stronger charge attractive forces near the device walls (\tab\ref{tab:1}). For instance, $E_\text{x,max}$ changes at $K=2$ and 20 by (-9.18\%, -13.98\%, -15.99\%, -15.94\%) and (275.71\%, 282.15\%, 284.52\%, 285.36\%) for ($d_\text{c}=0.25$, 0.50, 0.75, 1), respectively when $S$ varies from 4 to 16 (refer \tab\ref{tab:1}). Further, $E_\text{x,max}$ enhances with decreasing $d_\text{c}$ because reduction in the cross-flow area increases excess ions ($n^\ast$) as discussed earlier in the section \ref{sec:charge} and velocity enhancement in the channel, therefore, increases the induced electric field strength. The relative effect of $d_\text{c}$ on $E_\text{x,max}$ is maximum at higher $K$ and lower $S$ (\tab\ref{tab:1}). For instance, $E_\text{x,max}$ enhances by (84.34\%, 84.23\%, 84.24\%) and (133.99\%, 133.88\%, 133.48\%) for ($S=4$, 8, 16), respectively at $K=2$ and 20 with $d_\text{c}$ varies from 1 to 0.75; corresponding increment in the values of $E_\text{x,max}$ with decreasing $d_\text{c}$ from 1 to 0.50 are recorded as (234.80\%, 230.23\%, 242.61\%) and (657.41\%, 656.14\%, 651.12\%), respectively at $K=2$ and 20. Similarly, $E_\text{x,max}$ increases by (707.26\%, 689.26\%, 772.17\%) and (5280.51\%, 5251.95\%, 5145.75\%), at $K=2$ and 20, respectively with overall decreasing contraction $d_\text{c}$ from 1 to 0.25 ($0.25\le d_\text{c}\le 1$) (refer \tab\ref{tab:1}). 
\begin{figure}[h]
\centering\includegraphics[width=1\linewidth]{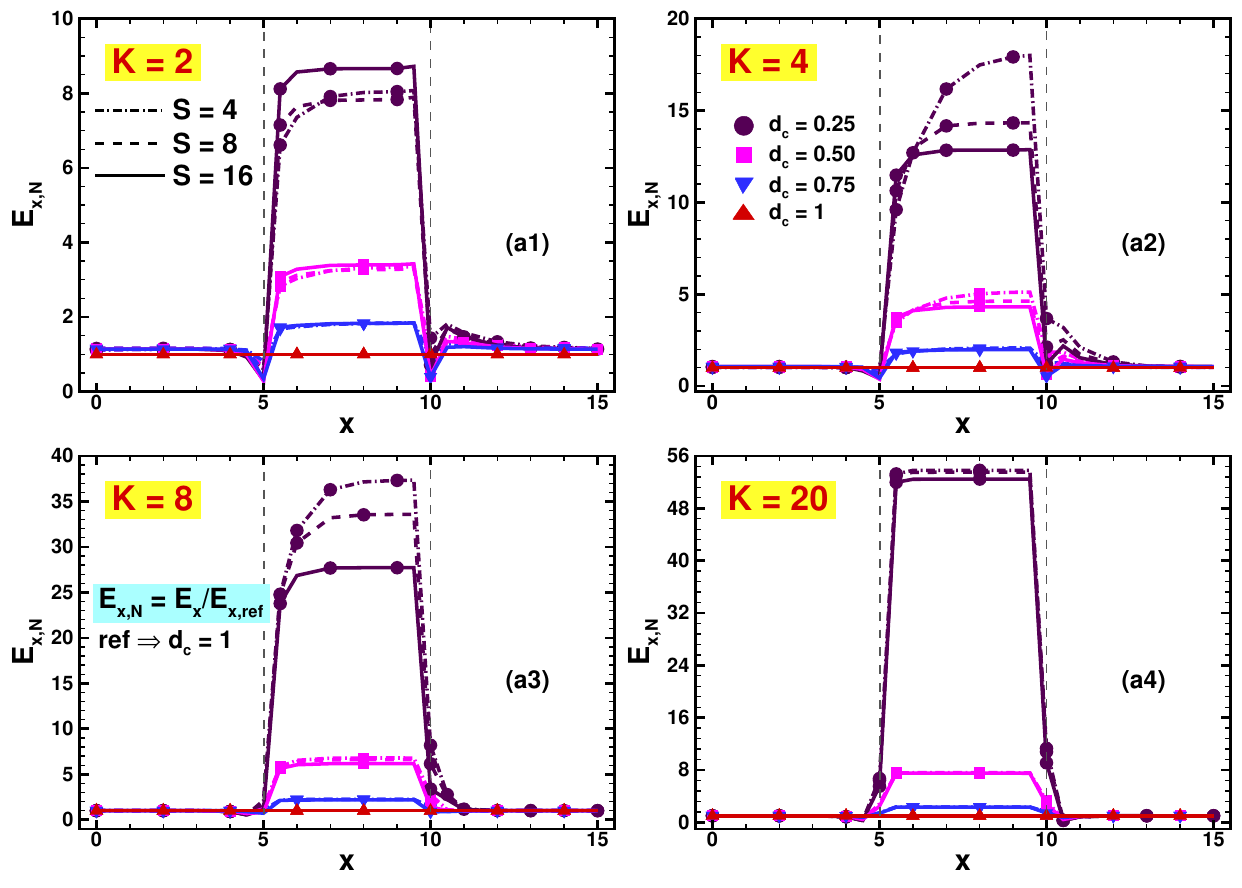}
\caption{Centreline profiles of normalized induced electric field strength ($E_\text{x,N}$) for $2\le K\le 20$, $4\le S\le 16$, and $0.25\le d_\text{c}\le 1$.}
\label{fig:4d}
\end{figure} 
\\\noindent Subsequently, for detailed investigation of induced electric field strength ($E_\text{x}$) variation on the centreline ($0\le x\le L$, 0) of device, it is normalized by reference case (at $d_\text{c}=1$). \fig\ref{fig:4d} represents centreline profiles of normalized induced electric field strength ($E_\text{x,N}$, \eqn\ref{eq:11a}) in the charged microchannel for $2\le K\le 20$, $4\le S\le 16$, and $0.25\le d_\text{c}\le 1$. Normalized induced electric field strength ($E_\text{x,N}$) variation along the length is similar qualitatively as $E_\text{x,n}$ (\fig\ref{fig:4c}). The $E_\text{x,N}$ is maximum in the contraction as compared to upstream and downstream sections of device (\fig\ref{fig:4d}). Maximum normalized induced electric field strength ($E_\text{x,N,max}$) increases with increasing $K$ (\fig\ref{fig:4d}) because $E_\text{x,max}$ variation with $d_\text{c}$ is greater at higher $K$, irrespective of $S$ (refer \tab\ref{tab:1}). Fro instance, maximum enhancement in $E_\text{x,N,max}$ is noted as 578.10\% (7.8926 to 53.5195) at $d_\text{c}=0.25$ and $S=8$ when $K$ changes from 2 to 20 (refer \fig\ref{fig:4d}). The $E_\text{x,N,max}$ enhances with decreasing $S$ but it has shown reverse trends at $K=2$ (\fig\ref{fig:4d}a1); the change of $E_\text{x,N,max}$ with $S$ is depicted maximum at $K=4$. For instance, $E_\text{x,N,max}$ reduces by 20.37\% (18.0018 to 14.3345) and 28.50\% (18.0018 to 12.8711) with increasing $S$ (from 4 to 8) and (from 4 to 16), respectively at $d_\text{c}=0.25$ and $K=4$ in the contraction section ($x=8$) of device (refer \fig\ref{fig:4d}a2). Further, $E_\text{x,N,max}$ increases with decreasing $d_\text{c}$ due to reduction in the constricted flow area of middle section of device (\fig\ref{fig:4d}). It is because more excess charge transport due to clustering of $n^\ast$ (refer section \ref{sec:charge}) and velocity enhancement, increases $E_\text{x}$ and hence normalized induced electric field strength. The relative impact of $d_\text{c}$ on $E_\text{x,N,max}$ is noted maximum at lower $S$ and higher $K$ due to maximum relative variation of $E_\text{x}$ for identical ranges of conditions (refer \tab\ref{tab:1}). 
%%
%For instance, $E_\text{x,N,max}$ maximally enhances by 5280.51\% (1 to 53.8051) when $d_\text{c}$ varies from 1 to 0.25 at $S=4$, $K=20$ (refer \fig\ref{fig:4d}a4).
%
\\\noindent The preceding discussion has depicted stronger dependence of electrostatic and ionic fields such as total electrical potential ($U$), excess ionic charge ($n^\ast$), and induced electric field strength ($E_\text{x}$) on governing parameters ($S$, $K$, $d_\text{c}$). The next section presents and analysis the corresponding influences on pressure ($P$).  
%
%---------------------------------
\subsection{Velocity ($\myvec{V}$) field}
\begin{figure}[!tb]
\centering\includegraphics[width=1\linewidth]{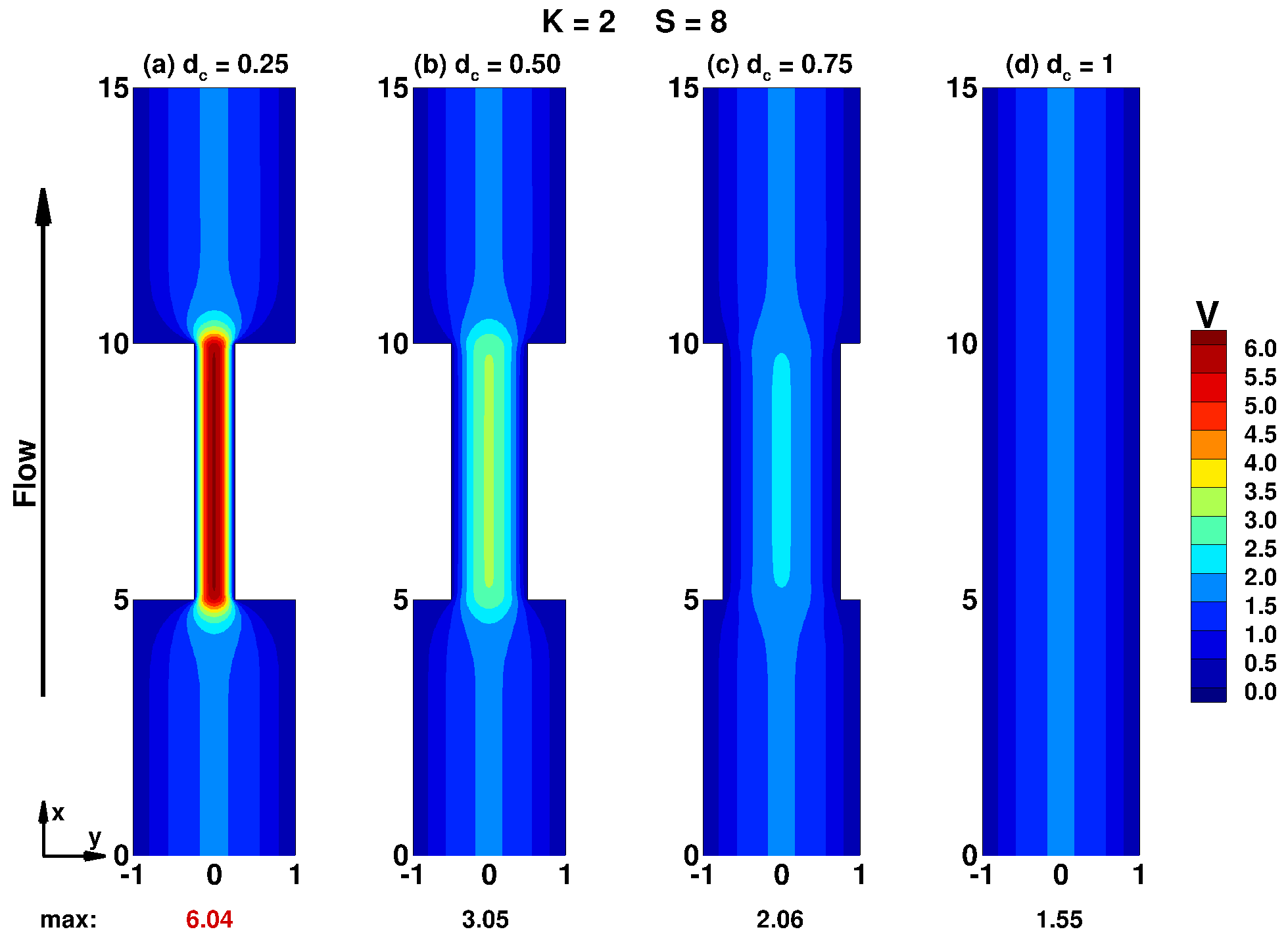}
\caption{Velocity ($\myvec{V}$) field distribution for $0.25\le d_\text{c}\le 1$ at $K=2$ and $S=8$.}
\label{fig:vc}
\end{figure}

\begin{figure}[htbp]
\centering\includegraphics[width=1\linewidth]{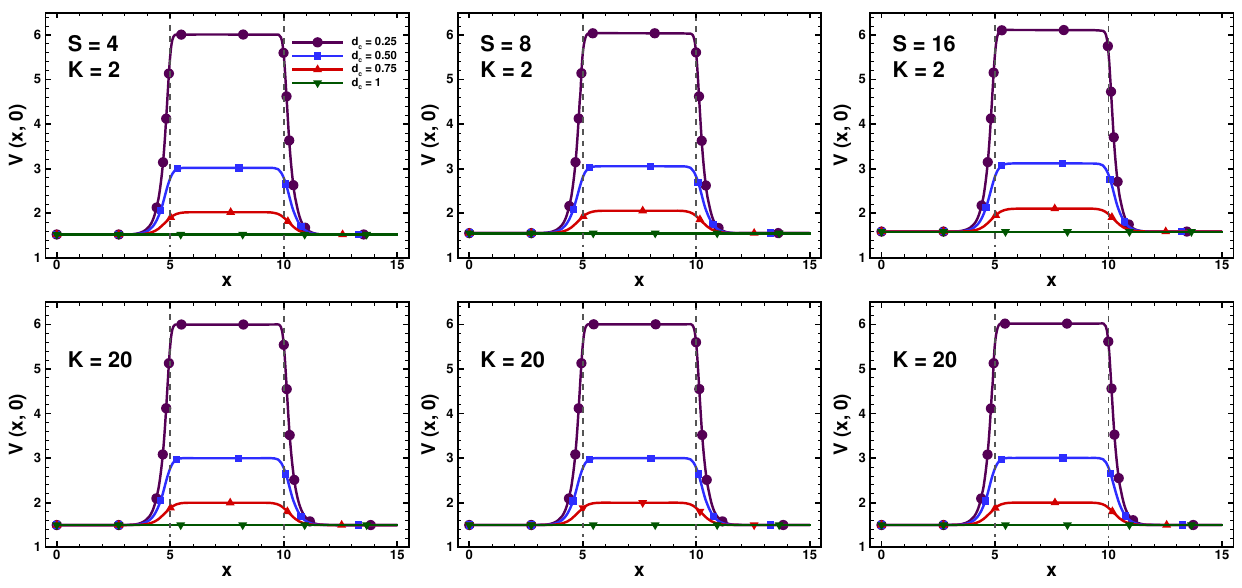}
\caption{Centreline profiles of velocity ($\myvec{V}$) field for $2\le K\le 20$, $4\le S\le 16$, and $0.25\le d_\text{c}\le 1$.}
\label{fig:v1}
\end{figure}

\begin{figure}[htbp]
\centering\includegraphics[width=1\linewidth]{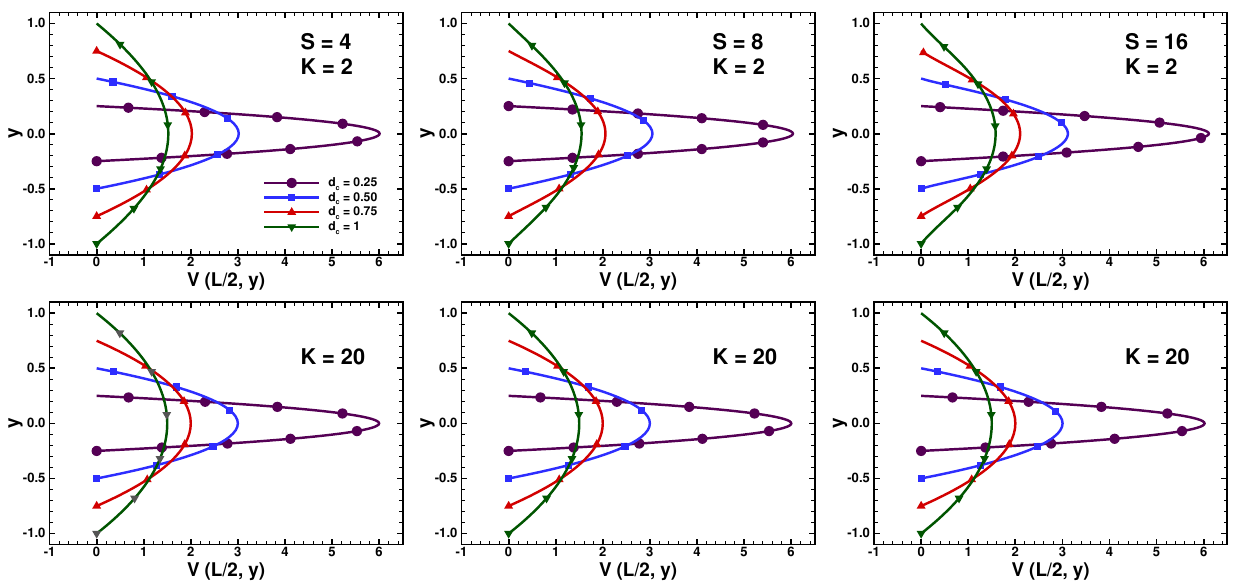}
\caption{Velocity ($\myvec{V}$) field variation over the vertical length (L/2, y) in the contraction for $2\le K\le 20$, $4\le S\le 16$, and $0.25\le d_\text{c}\le 1$.}
\label{fig:v2}
\end{figure}

%---------------------------------
%%%%%%%%%%
\label{sec:velocity}
%%%%%%%%%%%%%%%
%%
\noindent Velocity ($\myvec{V}$) field distribution in the considered microfluidic device for $0.25\le d_\text{c}\le 1$, $K=2$ and $S=8$ is depicted in the \fig\ref{fig:vc}. Contours for other ranges of condition ($2\le K\le 20$, $4\le S\le 16$, $0.25\le d_\text{c}\le 1$) are qualitatively similar, thus not presented here. Velocity ($\myvec{V}$) field increases from surface to centreline due to no-slip condition at the walls of microfluidic device. Therefore, the maximum value of $\myvec{V}$ is obtained at the centreline of device. In addition, $\myvec{V}$ enhances with decreasing $d_\text{c}$ due to reduction in the cross-section area at the fixed volumetric flow rate (i.e., $Q=A\myvec{V}=A_\text{c}\myvec{V}_\text{c}=$ constant). For instance, maximum value of $\myvec{V}$ varies from 1.55 to 6.04 with decreasing $d_\text{c}$ from 1 to 0.25 at $K=2$ and $S=8$ (\fig\ref{fig:vc}). However, overall maximum value of $\myvec{V}$ is obtained as 6.11 at $d_\text{c}=0.25$, $K=2$ and $S=16$ (\tab\ref{tab:vl}).
\\\noindent \fig\ref{fig:v1} shows the centreline (x, 0) profiles of velocity ($\myvec{V}$) field in the considered microfluidic device for $2\le K\le 20$, $4\le S\le 16$, and $0.25\le d_\text{c}\le 1$. The $\myvec{V}$ has shown the complex variation along the length of device, irrespective of $K$ and $S$. For instance, $\myvec{V}$ is constant in the upstream section followed by the steep enhance in the end of section. In contraction region, $\myvec{V}$ is constant throughout the section. Further, in downstream section, $\myvec{V}$ reduces in the starting of the section and becomes constant in the latter part of the section. Thus, the maximum velocity ($V_\text{max}$) is observed in the contraction than other sections of device. However, $\myvec{V}$ is constant throughout the channel at $d_\text{c}=1$, irrespective of $K$ and $S$. The $V_\text{max}$ has depicted the strong dependence on $d_\text{c}$ and insignificant dependence on $K$ and $S$. The increment in $V_\text{max}$ is noted with decreasing $d_\text{c}$. It is because at constant flow rate ($Q$), suddenly constricted flow area with reduction in $d_\text{c}$ increases $V_\text{max}$. For instance, $V_\text{max}$ for non-electroviscous ($S=0$ or $K=\infty$) flow is obtained as (1.5, 2, 3, 6) at ($d_\text{c}=1$, 0.75, 0.50, 0.25), respectively. The $V_\text{max}$ increases from (1.52 to 6.01), (1.55 to 6.04), and (1.58 to 6.11) for ($S=4$, 8, and 16), respectively with the variation of $d_\text{c}$ from 1 to 0.25 at $K=2$. Corresponding enhancement in $V_\text{max}$ is recorded as (1.50 to 6), (1.50 to 6.01), and (1.50 to 6.02) at $K=20$ (\tab\ref{tab:vl}). Further, slight increment in $V_\text{max}$ is observed with decreasing $K$ and increasing $S$ due to enhances electrostatic forces near the device walls.
\begin{table}
\centering
\caption{Maximum velocity ($V_\text{max}$) on the centreline of charged microfluidic device as function of governing parameters ($K$, $S$, and $d_\text{c}$).}\label{tab:vl}
\vspace{2mm}
\scalebox{1}
{
	\begin{tabular}{|r|r|r|r|r|r|}
		\hline
		$S$	&	$K$	&	\multicolumn{4}{c|}{$V_\text{max}$}	\\\cline{3-6}
		&		&	$d_\text{c}=0.25$	&	$d_\text{c}=0.50$	& $d_\text{c}=0.75$ &	$d_\text{c}=1$  \\\hline
		0   & 	$\infty$    & 	6	    & 3	        & 2          & 	1.5	 \\\hline
		4	&	2	&	6.0102	&	3.0183	&	2.0220	&	1.5209	\\
		&	20	&	6.0015	&	2.9997	&	1.9992	&	1.4991	\\\hline
		8	&	2	&	6.0399	&	3.0537	&	2.0561	&	1.5491	\\
		&	20	&	6.0061	&	3.0007	&	1.9996	&	1.4993	\\\hline
		16	&	2	&	6.1129	&	3.1173	&	2.1034	&	1.5818	\\
		&	20	&	6.0229	&	3.0047	&	2.0010	&	1.4999	\\\hline
	\end{tabular}
}
\end{table}
\\\noindent Subsequently, \fig\ref{fig:v2} represents the velocity ($\myvec{V}$) field variation over the vertical length (L/2, y) in the middle section of considered microfluidic device for $2\le K\le 20$, $4\le S\le 16$ and $0.25\le d_\text{c}\le 1$. The increment in $\myvec{V}$ is noted (0 to $V_\text{max}$) from walls to centreline of device. It is because no-slip condition is used at the devices walls. The $d_\text{c}$ has significantly affected $V_\text{max}$, irrespective of $K$ and $S$. The enhancement in $V_\text{max}$ is obtained with reducing $d_\text{c}$. However, relatively less increment is noted in $V_\text{max}$ with enhancing $S$ and decreasing $K$. It is because strong electrostatic forces reduce $V_\text{max}$ in the close vicinity of walls, thus, increases $V_\text{max}$ on the centreline of device.
%%
%
%---------------------------------
\subsection{Pressure ($P$) distribution}
\begin{figure}[h]
\centering\includegraphics[width=1\linewidth]{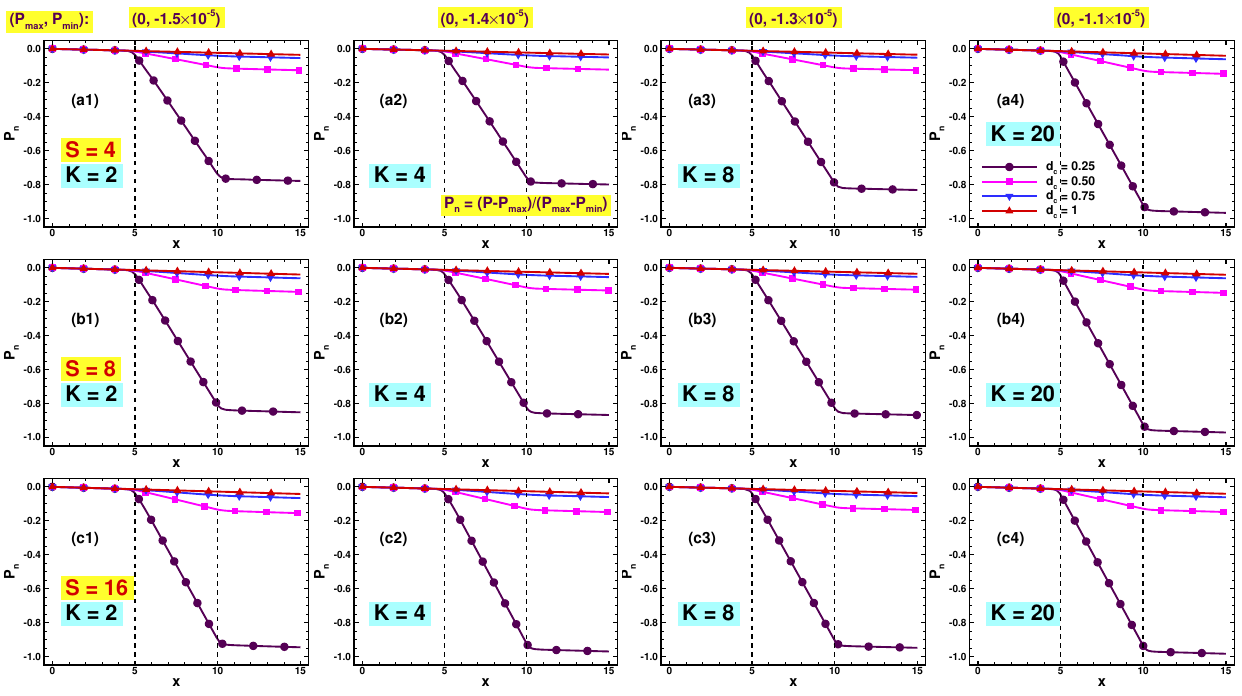}
\caption{Centreline profiles of normalized pressure ($P_\text{n}$) for $2\le K\le 20$, $4\le S\le 16$, and $0.25\le d_\text{c}\le 1$.}
\label{fig:5c}
\end{figure}

%---------------------------------
%%%%%%%%%%
\label{sec:pressure}
%%%%%%%%%%%%%%%
%
\noindent Normalized pressure is defined as $P_\text{n}=\frac{P-P_\text{max}}{P_\text{max}-P_\text{min}}$, here subscripts \textit{min} and \textit{max} denote minimum and maximum value of $P$ for each $K$. \fig\ref{fig:5c} depicts centreline profiles of normalized pressure ($P_\text{n}$) in the charged microchannel for $2\le K\le 20$, $4\le S\le 16$, and $0.25\le d_\text{c}\le 1$. The $P_\text{n}$ decreases along the length of microfluidic device (\fig\ref{fig:5c}) due to imposed additional hydrodynamic resistance on the fluid by increased $|\Delta U|$ along the length as we have discussed in the section \ref{sec:potential}. Normalized pressure gradient is maximum in the contraction section because of applied additional resistance by both charge attractive and hydrodynamic forces with sudden constricted flow area (\fig\ref{fig:5c}). Qualitative behavior of centreline $P_\text{n}$ profiles are same as $P$ with the literature \citep{davidson2007electroviscous,dhakar2022electroviscous}. Normalized pressure has shown complex dependency on $K$, $S$, and $d_\text{c}$. For instance, $|P_\text{n}|$ enhances maximally by 24.17\% (0.7782 to 0.9663) when $K$ varies from 2 to 20 at $S=4$ and $d_\text{c}=0.25$ (refer \fig\ref{fig:5c}a). The $P_\text{n}$ decreases with increasing $S$ (\fig\ref{fig:5c}). For instance, $|P_\text{n}|$ increases maximally by 22.89\% (0.1264 to 0.1553) when $S$ varies from 4 to 16 at $K=2$ and $d_\text{c}=0.50$ (refer \fig\ref{fig:5c}). Further, $|P_\text{n}|$ enhances with decreasing $d_\text{c}$ (\fig\ref{fig:5c}). The variation in the value of $P_\text{n}$ with $d_\text{c}$ is greater at higher $K$ and $S$. For instance, maximum increment in the value of $|P_\text{n}|$ is noted as 2526.38\% (0.0361 to 0.9484) when $d_\text{c}$ varies from 1 to 0.25 at $K=8$ and $S=16$ (refer \fig\ref{fig:5c}c3).
\\\noindent Subsequently, \tab\ref{tab:1} comprises the pressure drop ($|\Delta P|$) on centreline ($0\le x\le L$, 0) of device for $2\le K\le 20$, $4\le S\le 16$, and $0.25\le d_\text{c}\le 1$. The $\Delta P$ variation with $K$ and $S$ is similar as the literature \citep{davidson2007electroviscous,dhakar2022electroviscous,dhakar2023cfd} for a fixed $d_\text{c}=0.25$. Pressure drop ($|\Delta P|$) increases with decreasing $K$ or EDL thickening; the change in the values of $|\Delta P|$ with $K$ are maximum at higher $S$ (\tab\ref{tab:1}). For instance, pressure drop ($|\Delta P|$) reduces by (8.94\%, 14.35\%, 17.12\%, 15.57\%) and (23.73\%, 29.95\%, 31.80\%, 28.78\%) for ($d_\text{c}=0.25$, 0.50, 0.75, 1) at $S=4$ and 16 when $K$ varies from 2 to 20 (refer \tab\ref{tab:1}). The $|\Delta P|$ enhances with increasing $S$ due to additional resistance imposed by charge attractive forces with increasing $S$ (\tab\ref{tab:1}). The relative effect of $S$ on $|\Delta P|$ is maximum at thick EDL (lower $K$). For instance, increment in $|\Delta P|$ is recorded at $K=2$ and 20 as (21.48\%, 22.89\%, 21.82\%, 18.77\%) and (1.75\%, 0.51\%, 0.24\%, 0.18\%) for ($d_\text{c}=0.25$, 0.50, 0.75, 1), respectively when $S$ varies from 4 to 16 (refer \tab\ref{tab:1}). Further, $|\Delta P|$ increases with decreasing $d_\text{c}$ (\tab\ref{tab:1}) because enhancement in the additional hydrodynamic resistance on the fluid applied by increase total potential with reducing $d_\text{c}$ as discuss in the section \ref{sec:potential}. The variation in the value of $|\Delta P|$ with $d_\text{c}$ is maximum at higher $S$ and $K$. For instance, $\Delta P$ enhances for ($S=4$, 8, 16) by (53.54\%, 56.06\%, 57.48\%) and (50.72\%, 50.74\%, 50.82\%), respectively at $K=2$ and 20 when $d_\text{c}$ varies from 1 to 0.75; corresponding increment in the values of $|\Delta P|$ are recorded as (256.02\%, 258.54\%, 268.39\%) and (261.17\%, 261.42\%, 262.38\%) with decreasing $d_\text{c}$ from 1 to 0.50. Similarly, $\Delta P$ increases by (2091.78\%, 2050.67\%, 2141.79\%) and (2263.68\%, 2271.73\%, 2300.83\%), respectively at $K=2$ and 20 with overall decreasing contraction $d_\text{c}$ from 1 to 0.25 ($0.25\le d_\text{c}\le 1$) (refer \tab\ref{tab:1}).
\begin{figure}[h]
\centering\includegraphics[width=1\linewidth]{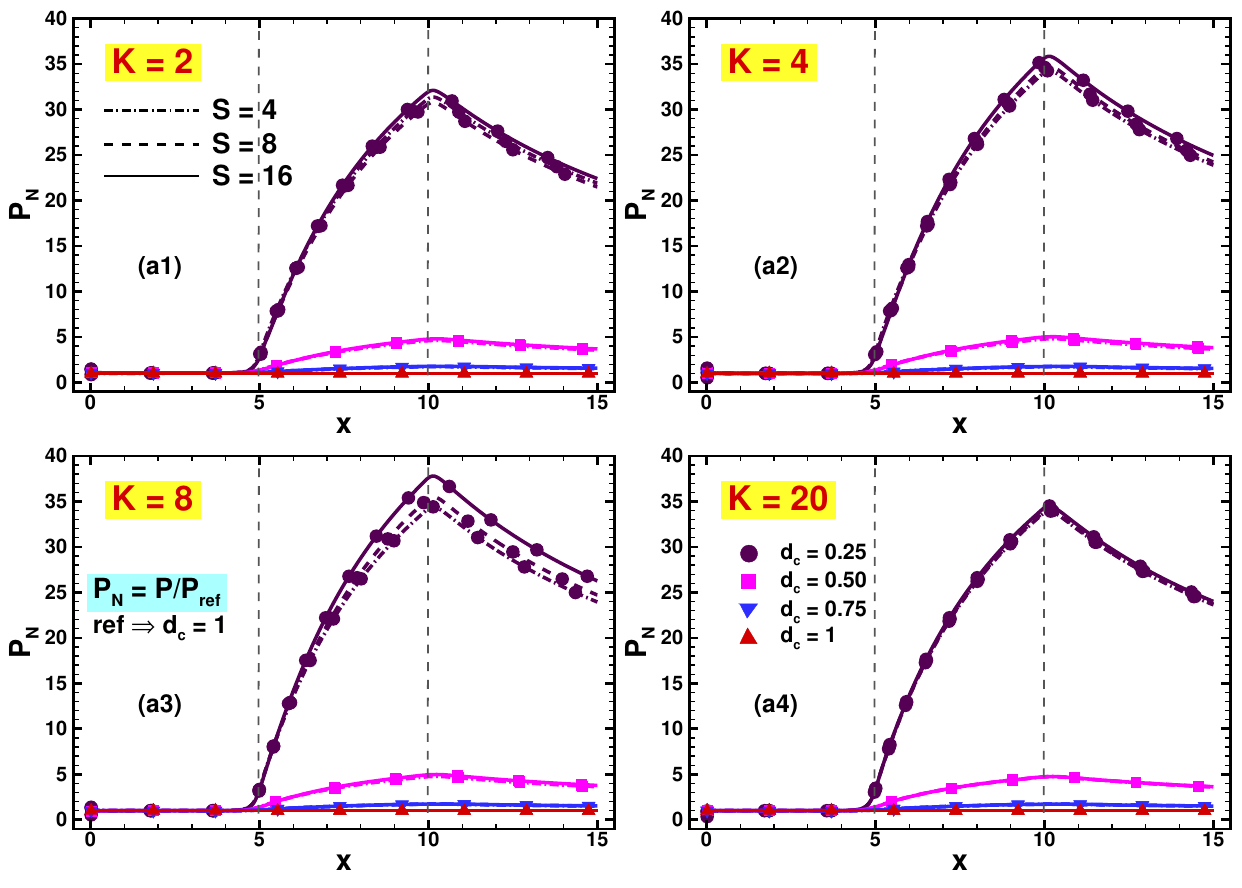}
\caption{Centreline profiles of normalized pressure ($P_\text{N}$) for $2\le K\le 20$, $4\le S\le 16$, and $0.25\le d_\text{c}\le 1$.}
\label{fig:5d}
\end{figure} 
\\\noindent Further, pressure ($P$) is normalized by reference case (at $d_\text{c}=1$) for detailed investigation of pressure variation along the centreline ($0\le x\le L$, 0) of microfluidic device. \fig\ref{fig:5d} depicts centreline profiles of normalized pressure ($P_\text{N}$, \eqn\ref{eq:11a}) in the charged microchannel for $2\le K\le 20$, $4\le S\le 16$, and $0.25\le d_\text{c}\le 1$. Normalized pressure ($P_\text{N}$) qualitative variation along the length of microfluidic device is same as $U_\text{N}$ discussed in the section \ref{sec:potential} (refer \fig\ref{fig:2d}). Normalized pressure increases with increasing $K$ (\fig\ref{fig:5d}) because the variation of $P$ with $d_\text{c}$ is greater at higher $K$, irrespective of $S$ (\tab\ref{tab:1}). For instance, maximum increment in the value of $P_\text{N}$ is noted as 10.28\% (21.5067 to 23.7173) when $K$ changes from 2 to 20 at $d_\text{c}=0.25$ and $S=8$ (refer \fig\ref{fig:5d}). Normalized pressure increases with decreasing $S$; variation in the value of $P_\text{N}$ with $S$ is maximum at $K=8$ (\fig\ref{fig:5d}). For instance, $P_\text{N}$ enhances by $3.18\%$ ($23.9478$ to $24.7101$) and $9.67\%$ ($23.9478$ to $26.2638$) with increasing $S$ (from 4 to 8) and (from 4 to 16), respectively at $d_\text{c}=0.25$ and $K=8$ in the end location ($x=10$) of contraction section (refer \fig\ref{fig:5d}a3). Further, $P_\text{N}$ increases with decreasing $d_\text{c}$ due to reducing area of cross-section in the contraction section. It is because enhances convective transport of excess ions in the device, decreases total potential (refer section \ref{sec:potential}). Therefore, it is imposed a additional hydrodynamic force on the fluid and increases normalized pressure (\fig\ref{fig:5d}). The relative impact of $d_\text{c}$ on $P_\text{N}$ is maximum at higher $K$ and lower $S$ because EDLs do not occupy the greater fraction of microfluidic device (\fig\ref{fig:5d}). 
%%
%For instance, $P_\text{N}$ increases maximally by 2300.83\% (1 to 24.0083) with decreasing contraction $0.25\le d_\text{c}\le 1$ at $S=16$, $K=20$ (refer \fig\ref{fig:5d}a4).
%%
%%%
%%
%---------------------------------
\subsection{Electroviscous correction factor ($Y$)}
\begin{figure}[htbp]
\centering
\subfigure[]{\includegraphics[width=0.49\linewidth]{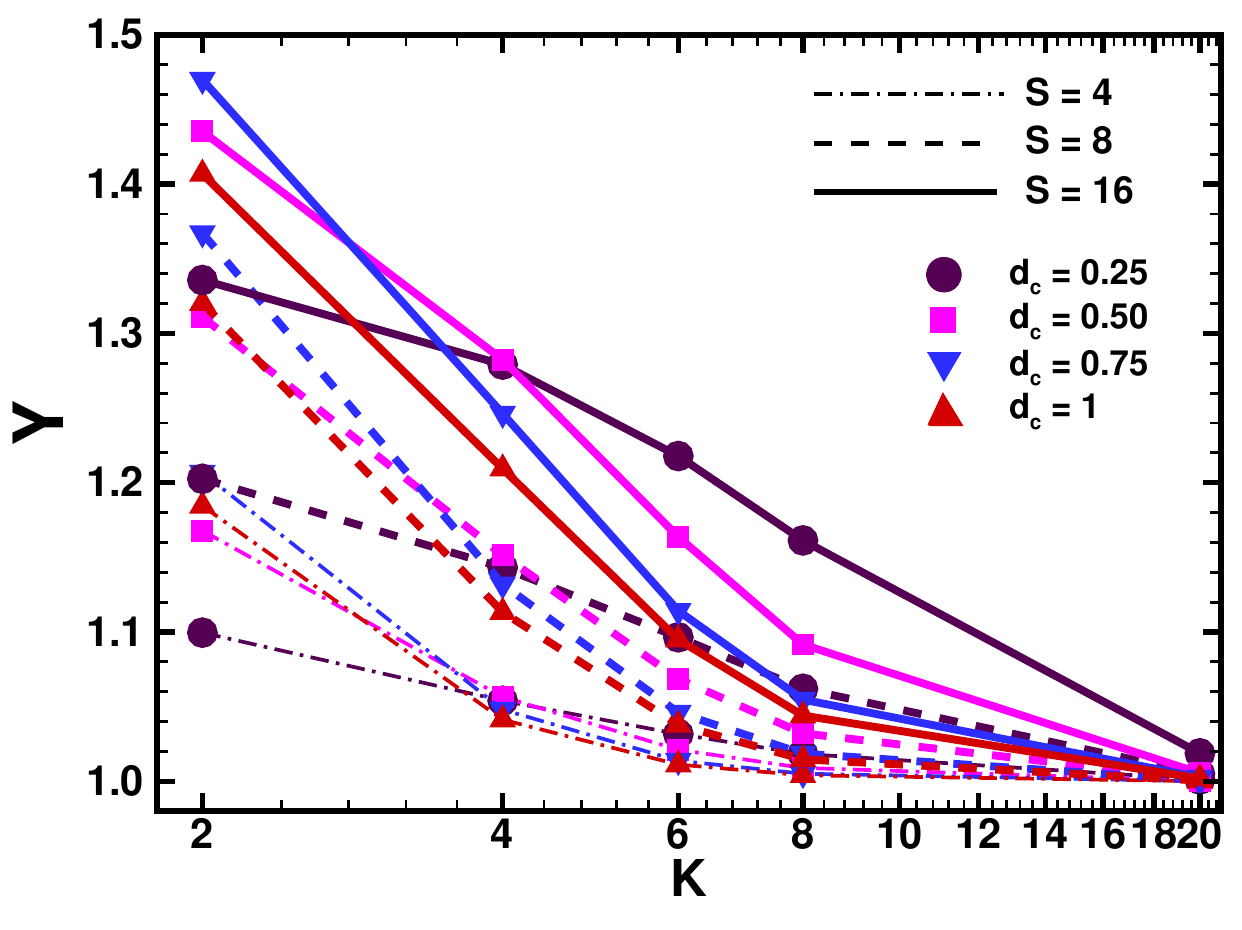}}
\subfigure[]{\includegraphics[width=0.49\linewidth]{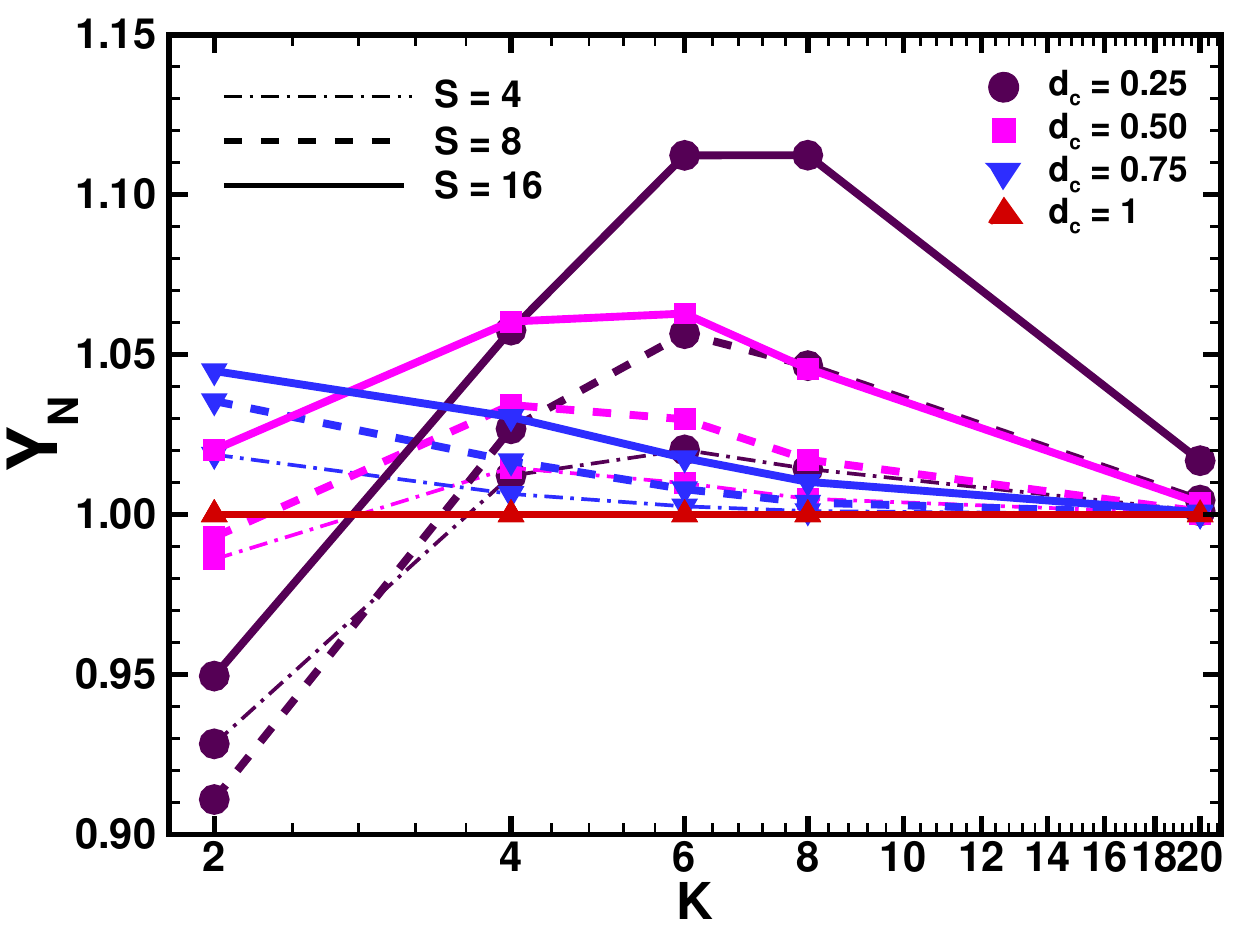}}
\caption{(a) Electroviscous correction factor ($Y$), (b) Normalized electroviscous correction factor ($Y_\text{N}$).}
\label{fig:6}
\end{figure}

%%%%---------------------------------
\label{sec:ECF}
%%%%
\noindent In the electrokinetic flows, the convective transport of excess ions ($n^\ast$) develops a induced electric field strength ($E_\text{x}$) that imposes a additional hydrodynamic resistance on the fluid flow in the microfluidic device. This extra hydrodynamic resistance manifests a pressure drop ($\Delta P$) that is greater than the pressure drop ($\Delta P_0$) excluding electrical body forces ($S=0$) at the fixed volumetric flow rate ($Q$). It is generally described as the apparent viscosity ($\mu_{\text{eff}}$) that is the viscosity needed to acquires pressure drop ($\Delta P$) in the absence electrical body forces ($S=0$). Further, it is quantified as the `electroviscous effect' (EVE) \citep{davidson2007electroviscous,bharti2008steady,dhakar2022electroviscous}. For low Reynolds number steady laminar PDF flow, the non-linear advection term in the momentum equation (\eqn\ref{eq:5}) becomes negligible. The relative enhancement in the pressure drop ($\Delta P/\Delta P_0$) relates the relative enhancement in the viscosity ($\mu_\text{eff}/\mu$) of fluid, under otherwise identical conditions. Thus, \textit{electroviscous correction factor} ($Y$) is expressed as follows. 
\begin{gather}
Y=\frac{\mu_{\text{eff}}}{\mu}=\frac{\Delta P}{\Delta P_{\text{0}}}
\label{eq:27}
\end{gather}
where $\mu$ is viscosity of liquid yielding the pressure drop ($\Delta P_0$).
\\\noindent \fig\ref{fig:6}a represents electroviscous correction factor ($Y$) as function of $2\le K\le 20$, $4\le S\le 16$, and $0.25\le d_\text{c}\le 1$. Electroviscous effects are stronger when $Y>1$ and absent when $Y=1$. The factor ($Y$) shows significant dependency on $K$ and $S$, irrespective of $d_\text{c}$. The $Y$ depicts the similar behavior with $K$ and $S$ as the literature \citep{davidson2007electroviscous,dhakar2022electroviscous} for a fixed $d_\text{c}=0.25$. The correction factor ($Y$) decreases with increasing $K$ (EDL thinning) and increases with increasing $S$ for a fixed $d_\text{c}$ (\fig\ref{fig:6}a). For instance, $Y$ enhances by 31.80\% (at $S=16$, $d_\text{c}=0.75$) and 22.89\% (at $K=2$, $d_\text{c}=0.50$) with decrement of $K$ (from 20 to 2) and increment of $S$ (from 4 to 16), respectively (refer \fig\ref{fig:6}a). 
\\\noindent The factor ($Y$) has shown complex dependency on $d_\text{c}$. At $K>4$, $Y$ increases with decreasing $d_\text{c}$ for $4\le S\le 16$ (\fig\ref{fig:6}a). It is because reduction in the contraction area enhances velocity and hence pressure drop (i.e., $\Delta P\propto \myvec{V}$ for a channel flow by \textit{Hagen-Poiseuille} relation), therefore, $Y$ increases from \eqn(\ref{eq:27}). However, at $K=4$, the variation of $Y$ with $d_\text{c}$ has shown the cross-over point (\fig\ref{fig:6}a). At $K<4$, $Y$ is obtained maximum (at $d_\text{c}=0.75$) and minimum (at $d_\text{c}=0.25$) for $4\le S\le 16$ (\fig\ref{fig:6}a). Further, at $K=2$, $Y$ increases with $d_\text{c}$ varies from 0.25 followed by 0.50, 1, and 0.75, respectively for $S=4, 8$; corresponding $Y$ enhances with $d_\text{c}$ changes from $0.25$ followed by 1, 0.50, and 0.75, respectively for $S=16$. It is because EDLs start to overlap at lower $K\le4$, therefore, such complex behavior is obtained over given ranges of conditions. Overall increment in $Y$ is noted as 46.99\% (at $K=2$, $S=16$, $d_\text{c}=0.75$), relative to non-EVF ($S=0$) (refer \fig\ref{fig:6}a). 
% 
%%%
%
\\\noindent Further, the complex variation of factor ($Y$) with governing parameters ($K$, $S$, $d_\text{c}$) is understood in detailed by normalizing $Y$ with uniform ($d_\text{c}=1$) geometry case. \fig\ref{fig:6}b depicts the normalized electroviscous correction factor ($Y_\text{N}$, \eqn\ref{eq:11a}) as a function of dimensionless parameters ($2\le K\le 20$, $4\le S\le 16$, $0.25\le d_\text{c}\le 1$). The $Y_\text{N}$ is equal to unity for $d_\text{c}=1$ due to normalizing with that case (\fig\ref{fig:6}b). Normalized factor ($Y_\text{N}$) shows complex dependency on $K$, $S$, and $d_\text{c}$. The $Y_\text{N}$ increases with increasing $S$, irrespective of the $K$ and $d_\text{c}$ (\fig\ref{fig:6}b). The $Y_\text{N}$ decreases with increasing $K$ for $d_\text{c}=0.75$ and $4\le S\le 16$; $Y_\text{N}$ increases for $2\le K\le4$ and decreases for $4\le K\le 20$, respectively at $d_\text{c}=0.50$ and $4\le S\le 16$. However, $Y_\text{N}$ increases for $2\le K\le 6$ and reduces for $6\le K\le 20$ at $d_\text{c}=0.25$ and $4\le S\le 16$ (\fig\ref{fig:6}b). For instance, $Y_\text{N}$ maximally increases by 10.28\% ($S=8$, $d_\text{c}=0.25$) when $K$ varies from 2 to 20. Normalized correction factor has also shown complex variation with $d_\text{c}$ similar as $Y$ for given ranges of $2\le K\le 20$ and $4\le S\le 16$ (refer \fig\ref{fig:6}a). Normalized factor increases with decreasing $d_\text{c}$ for $4\le K\le 20$ (\fig\ref{fig:6}b) due to enhancement in $Y$ with reducing the constricted flow area in the middle section, under otherwise identical conditions (refer \fig\ref{fig:6}a). For instance, $Y_\text{N}$ maximally enhances by 11.24\% when $d_\text{c}$ varies from 1 to 0.25 at $S=16$ and $K=8$ (refer \fig\ref{fig:6}b).
\\\noindent Electroviscous correction factor ($Y$) functional dependence on the dimensionless parameters ($K$, $S$, $d_\text{c}$) is given below. 
\begin{gather}
Y = A_{1} + (A_{2}+A_4X)X + (A_{3} + A_{5}\gamma)\gamma + A_{6}X\gamma
\label{eq:32}
\\
\text{where}\qquad 
A_{\text{i}} = \sum_{{j}=1}^4 N_{\text{ij}} {d_\text{c}}^{({j}-1)};\qquad X=K^{-1};\qquad \gamma = S^{-1}; \qquad 1\le i \le 6 \nonumber
\end{gather}
where, the correlation coefficients ($N_{\text{ij}}$) are statistically obtained as 
\begin{gather*}
N = \begin{bmatrix}
	%M11 & M12 & M13 & M14 \\
	%M21 & M22 & M23 & M24 \\
	%M31 & M32 & M33 & M34\\
	%M41 & M42 & M43 & M44\\ 
	%M51 & M52 & M53 & M54\\ 
	%M61 & M62 & M63 & M64\\ 
	1.2228	&	0.7714	&	-2.6779	&	-1.3376	&	6.1330	&	0.0088	\\
	-0.8179	&	1.5979	&	6.4765	&	-2.3153	&	-15.8020	&	0.1167	\\
	0.6728	&	-6.3056	&	-1.3392	&	12.7820	&	2.4928	&	-0.1008	\\
	-0.1397	&	4.0203	&	-1.9083	&	-8.6731	&	5.3792	&	0.0139	
\end{bmatrix}^{T} 
\end{gather*}
by performing the non-linear regression analysis using the DataFit (trail version) with ($\delta_{\text{min}}$, $\delta_{\text{max}}$, $\delta_{\text{avg}}$, $R^2$) as (-3.94\%, 3.93\%, -0.04\%, 97.21\%) for given ranges of conditions.
%
%
%
%---------------------------------
\subsection{Pseudo-analytical model}
%%%%%%
\noindent The flow characteristics for electrolyte liquid flow through charged slit microfluidic device have been calculated numerically as discussed in the above sections. However, it can be calculate analytically for wide ranges of parameters and broad microfluidic applications for design aspects. Previous studies have proposed simpler analytical model to calculate the pressure drop ($\Delta P$) in the symmetrically ($S_\text{r}=1$) charged device for no-slip, charge-dependent slip flow and in the asymmetrically ($S_\text{r}\neq1$) charged microfluidic device with contraction-expansion ($d_\text{c}=0.25$) cross-section \citep{davidson2007electroviscous,bharti2008steady,dhakar2022electroviscous,dhakar2023cfd}. Analytically estimated pressure drop from these studies overpredicts the numerically estimated pressure drop maximally by $5-10\%$. 
\\\noindent In this study, earlier analytical model \citep{davidson2007electroviscous,bharti2008steady,dhakar2022electroviscous,dhakar2023cfd} is extended for electrolyte liquid flow through charged slit microfluidic device for variable contraction ratio ($0.25\le d_\text{c}\le 1$). The pressure drop ($\Delta P$) for laminar steady incompressible Newtonian fully developed flow through slit microfluidic device of length ($L$) for non-EVF ($S=0$) is computed  by summing up the individual uniform (upstream, contraction, downstream) sections pressure drop ($\Delta P_\text{u}$, $\Delta P_\text{c}$, $\Delta P_\text{d}$) by standard \textit{Hagen-Poiseuille} equation and excess pressure drop ($\Delta P_\text{e}$) because of sudden contraction/expansion is predicted by pressure drop through thin orifices ($d_\text{c}<<1$)
\citep{davidson2007electroviscous,bharti2008steady,sisavath2002creeping,pimenta2020viscous}. It is expressed as follows  
%---------------------------------
%%
%
\begin{gather}
\Delta P_{\text{00,m}}=\left(\sum_{i=u,c,d}\Delta P_{\text{00,i}}\right) +\Delta P_{\text{00,e}}
% (\Delta P_{\text{u}}+\Delta P_{\text{c}}+\Delta P_{\text{d}})+\Delta P_{\text{e}}
\label{eq:33}
\end{gather}
where, the subscripts $'00'$, $u$, $c$, and $d$ denote non-EVF ($S=0$ or $K=\infty$) with fixed contraction ($d_\text{c}=0.25$), upstream, contraction, and downstream sections, respectively and these sections individually represent slit microchannel of uniform cross-section. The pressure terms are expressed as follows.
\begin{gather}
\Delta P_{\text{0,i}} = \left(\frac{3}{Re}\right){\frac{\Delta L_{\text{i}}}{d_\text{i}^3}};\qquad 
\Delta P_{\text{0,e}} =\frac{16}{\pi d_{\text{c}}^2Re}; \qquad \text{where} \qquad d_\text{i}=\frac{W_\text{i}}{W}
\nonumber
\end{gather}
where $d_{\text{c}}$ is the contraction ratio. The lengths ($L_\text{u}$, $L_\text{d}$, and $L_\text{c}$) are scaled width $W$ and Reynolds number ($Re$) is defined in the \eqn(\ref{eq:1}).
\\\noindent Subsequently, \eqn(\ref{eq:33}) is extended, and generalized pressure drop for variable contraction ($0.25\le d_\text{c}\le1$) in absence of electric field ($S=0$) is expressed as follows.
\begin{gather}
\Delta P_{\text{0,m}}= \frac{3}{Re}\left[L_\text{u} + \frac{\alpha L_\text{c}}{d_\text{c}^3} + L_\text{d} + \frac{16\alpha}{3\pi d_{\text{c}}^2}\right]
\label{eq:38}
\end{gather}
The influence of contraction variation ($0.25\le d_\text{c}\le1$) on the pressure drop ($\Delta P_{00,\text{m}}$) is accounted by correction coefficient ($\alpha$, \eqn\ref{eq:38}) and it is expressed as follows.
\begin{gather}
\alpha = C_1+C_2d_\text{c}+C_3d_\text{c}^2
\label{eq:39}
\end{gather}
The correlation coefficients ($C_\text{i}$, \eqn\ref{eq:39}) are statistically obtained as $C_1=1.022$, $C_2=-0.0635$, and $C_2=-0.2148$ by performing the non-linear regression analysis with ($\delta_{\text{min}}$,  $\delta_{\text{max}}$, $\delta_{\text{avg}}$, $R^2$) as (-0.77\%, 0.68\%, -0.01\% ,99.99) for given ranges of conditions.
\\
\eqns(\ref{eq:33}) and (\ref{eq:38}) expressed a generalized simpler analytical model for the low Reynolds number ($Re=0.01$) flow of liquid through a slit microfluidic device  at fixed $d_\text{c}$ ($=0.25$) and variable $d_\text{c}$, respectively excluding electric field ($S=0$) and it is further extended considering both electroviscous ($S>0$) and contraction variation ($0.25\le d_\text{c}\le1$) effects is given below.
\begin{gather}
\Delta P_{\text{m}}
= \Gamma_\text{evcv} \Delta P_\text{0,m}
= \Gamma_\text{evcv} \left[\frac{3}{Re}\left(L_\text{u} + \frac{\alpha L_\text{c}}{d_\text{c}^3} + L_\text{d} + \frac{16\alpha}{3\pi d_{\text{c}}^2}\right)\right]
\label{eq:40}
\end{gather}
where subscript `evcv` denotes electroviscous and contraction variation (EVCV) effects.
\\\noindent The influence of both contraction variation ($0.25\le d_\text{c}\le1$) and electroviscous ($S>0$) effects on the pressure drop ($\Delta P_{0,\text{m}}$) is accounted by correction coefficient ($\Gamma_\text{evcv}$, \eqn\ref{eq:40}) and it is expressed as follows.
\begin{gather}
\Gamma_\text{evcv} = A_1 + (A_2+A_4X)X + (A_3+ A_5\gamma)\gamma + A_6 X^{0.5}\gamma
\nonumber
\\\text{where}\qquad 
A_{\text{i}} = \sum_{{j}=1}^4 N_{\text{ij}} {d_\text{c}}^{({j}-1)};\quad X=K^{-1}; \quad \gamma=S^{-1}; \quad 1\le i\le 6 
\label{eq:41}
\end{gather}
The correlation coefficients ($N_\text{ij}$, \eqn\ref{eq:41}) are statistically obtained as
\begin{gather*}
N = \begin{bmatrix}
	%M11 & M12 & M13 & M14 & M15 \\
	%M21 & M22 & M23 & M24 & M25 \\
	%M31 & M32 & M33 & M34 & M45 \\
	%M41 & M42 & M43 & M44 & M45 \\ 
	%M51 & M52 & M53 & M54 & M55 \\
	%M61 & M62 & M63 & M64 & M65 \\ 
	%M71 & M72 & M73 & M74 & M75 \\
	1.2444	&	-0.7322	&	0.6237	&	-0.153	\\
	1.0135	&	4.7458	&	-9.0319	&	4.4003	\\
	-2.8434	&	4.2179	&	0.6104	&	-2.1758	\\
	-1.4557	&	-3.8409	&	14.1	&	-8.8549	\\
	7.1203	&	-2.7138	&	-8.8752	&	6.9885	\\
	-0.6105	&	-7.8852	&	6.8532	&	-0.9721
\end{bmatrix} 
\end{gather*}
by performing the non-linear regression analysis with ($\delta_{\text{min}}$, $\delta_{\text{max}}$, $\delta_{\text{avg}}$, $R^2$) as (-4.10\%, 2.90\%, -0.05\%, 98.60\%) for given ranges of conditions.
\\\noindent The above discussed simple analytical model (\eqns\ref{eq:38} and \ref{eq:40}) for low Reynolds number ($Re=0.01$) electrolyte liquid flow through slit microfluidic device considering variable contraction ($0.25\le d_\text{c}\le 1$) is further extended to calculate the electroviscous correction factor is expressed as follows.
\begin{gather}
Y_\text{m}=\frac{{\Delta P}_\text{m}}{\Delta P_{\text{0,m}}}
\label{eq:28}
\end{gather}
\begin{figure}[htbp]
\centering
\subfigure[pressure drop] {\includegraphics[width=0.49\linewidth]{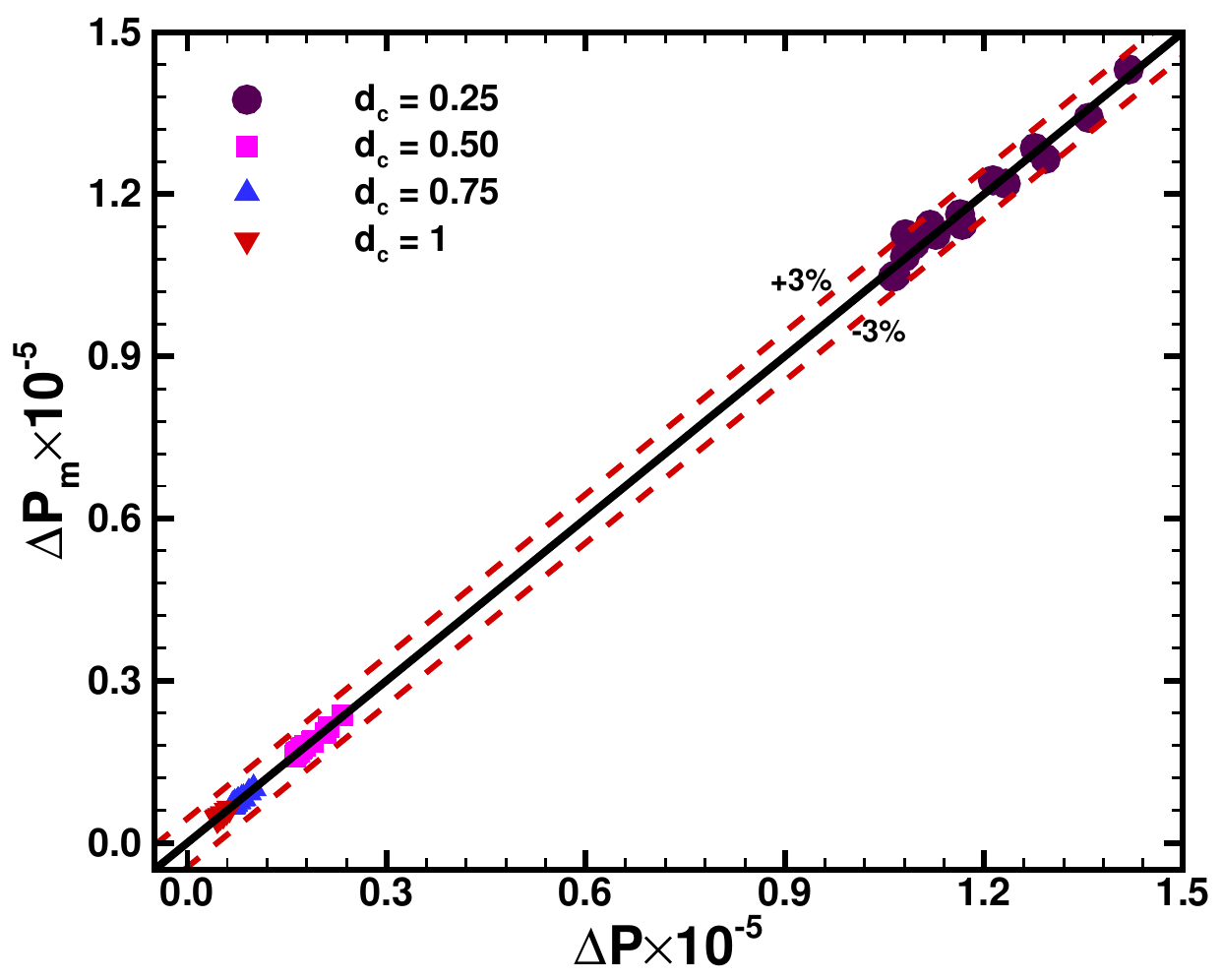}}
\subfigure[electroviscous correction factor] {\includegraphics[width=0.49\linewidth]{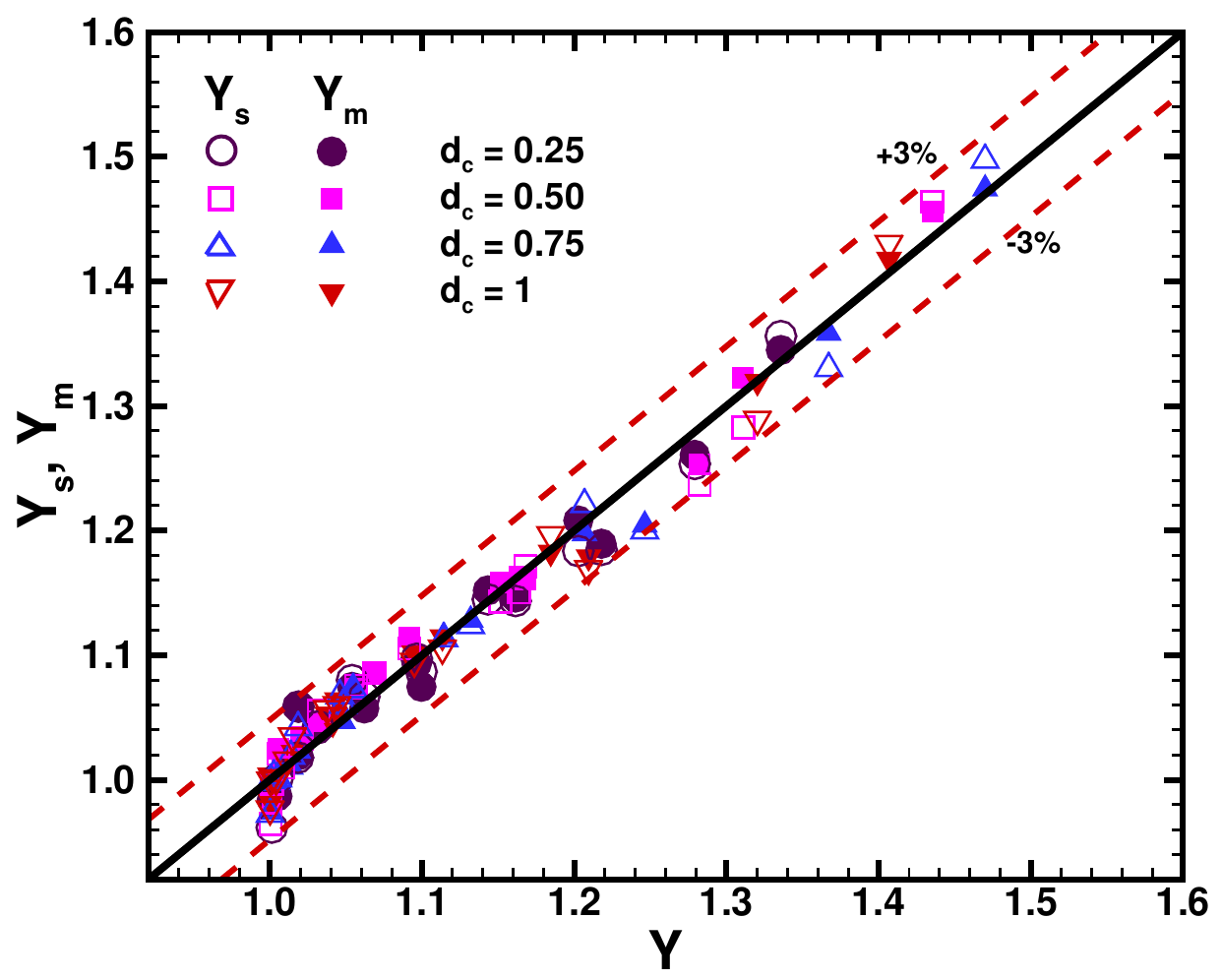}}
\caption{Parity chart for (a) numerically ($\Delta P$) and mathematically (${\Delta P}_\text{m}$, \eqn\ref{eq:40}) obtained pressure drop, (b) numerically ($Y$) and mathematically ($Y_\text{s}$, \eqn\ref{eq:32}; $Y_\text{m}$, \eqn\ref{eq:28}) obtained electroviscous correction factor for $K$, $S$ and $d_\text{c}$.}
\label{fig:7}
\end{figure} 
%
%%%
\noindent \figs\ref{fig:7}a and b depict parity charts for present numerical approach and simpler predictive model obtained pressure drop ($\Delta P$ vs $\Delta P_\text{m}$) and electroviscous correction factor ($Y$ vs $Y_\text{m}$) for $K$, $S$, and $d_\text{c}$. The simple analytical model overpredicts both pressure drop and electroviscous correction factor $\pm3\%$ of the numerical results. The difference between numerical and predicted results is smaller with decreasing surface charge density ($S$) and increasing inverse Debye length ($K$) or thinning of the EDL.  
%
%
%--------------------------------------
\section{Concluding remarks}
%--------------------------------------
%
\noindent In this work, contraction variation and electroviscous effects have investigated in the symmetric (1:1) electrolyte liquid flow through charged slit microfluidic device at low Reynolds number ($Re=10^{-2}$). A finite element method (FEM) is used to solve the mathematical model consisting of Poisson's, Nernst-Planck, and Navier-Stokes equations numerically to obtain the flow fields such as total electrical potential ($U$), ionic charge ($n_\pm$), excess charge ($n^\ast$), induced electric field strength ($E_\text{x}$), velocity ($\myvec{V}$) and pressure ($P$) fields for broader ranges of dimensionless governing parameters ($2\le K\le 20$, $4\le S\le 16$, $0.25\le d_\text{c}\le 1$).
\\\noindent Flow characteristics in the microfluidic device have shown complex dependency on the governing parameters ($K$, $S$, $d_\text{c}$). Total potential and pressure drop maximally change by 1785.58\% (0.2118 to 3.9929) and 2300.83\% (0.0450 to 1.0815), respectively with reducing contraction $d_\text{c}$ from 1 to 0.25, over the ranges of conditions. The electroviscous correction factor (i.e., ratio of apparent to physical viscosity) maximally increases by 11.24\% (at $K=8$, $S=16$), 31.80\% (at $S=16$, $d_\text{c}=0.75$), and 22.89\% (at $K=2$, $d_\text{c}=0.5$) when $d_\text{c}$ varies from 1 to 0.25 ($0.25\le d_\text{c}\le 1$), $K$ changes from 20 to 2 ($2\le K\le 20$), and $S$ varies from 4 to 16 ($4\le S\le 16$), respectively. Further, overall increment in electroviscous correction factor is noted as 46.99\% (at $K=2$, $S=16$, $d_\text{c}=0.75$), relative to non-EVE ($S=0$). Thus, increment in the contraction enhances the electroviscous effects in the microfluidic devices. However, it is worth noted that maximum $Y$ is obtained at $d_\text{c}=0.75$ than other $d_\text{c}$, for given ranges of $2\le K\le 20$ and $4\le S\le 16$.
\\\noindent A simple analytical model is developed to predict the pressure drop and electroviscous correction factor for broad ranges of flow governing parameters. This analytical model results estimate the pressure drop within $\pm$2-3\% to the present numerical results. The difference between the analytical and present numerical results reduces with decreasing $S$ and increasing $K$ or thinning of EDL. The present numerical results enable the use of these results to design essential and reliable microfluidic devices for various microfluidic applications. 
%
%
%\begin{comment}
%%%%%%%%%%%%%%%%%%%%%%%%%%%%%%%%%%%%%%%%%%
\section*{Declaration of Competing Interest}
%%%%%%%%%%%%%%%%%%%%%%%%%%%%%%%%%%%%%%%%%%
\noindent 
% All authors declare that they have no conflict of interest. 
The authors declare that they have no known competing financial interests or personal relationships that could have appeared to influence the work reported in this paper.
%
%The  authors  certify  that  they  have  NO  affiliations  with  or  involvement  in  any  organization or entity with any financial interest (such as honoraria; educational grants; participation in speakers’ bureaus; membership, employment, consultancies, stock ownership, or other equity interest; and expert testimony or patent-licensing arrangements), or non-financial interest (such as personal or professional relationships, affiliations, knowledge or beliefs) in the subject matter or materials discussed in this manuscript.
%%%%%%%%%%%%%%%%%%%%%%%%%%%%%%%%%%%%%%%%%%
%
\section*{Acknowledgements}
\noindent 
The authors acknowledge the infrastructural, computing resources, and software license support from the Indian Institute of Technology Roorkee. 
JD is thankful to the Department of Higher Education, Ministry of Education (MoE), Government of India (GoI) for the providence of research fellowship. 
%RPB would like to acknowledge Science and Engineering Research Board (SERB), Department of Science and Technology (DST), Government of India (GoI) for the providence of the MATRICS grant (File no. MTR/2019/001598). 
%
%--------------- Nomenclature
\begin{spacing}{1.5}
\input{Nomenclature.tex}

\renewcommand{\nompreamble}{\vspace{1em}\fontsize{10}{8pt}\selectfont}
%{\printnomenclature[5em]}
\printnomenclature
\end{spacing}
%
%\end{comment}
%\printglossaries 
%
%%%%%%%%%%%%%
\clearpage\appendix
\section{Additional information}
\setcounter{figure}{0}
\begin{figure}[h]
\centering\includegraphics[width=0.9\linewidth]{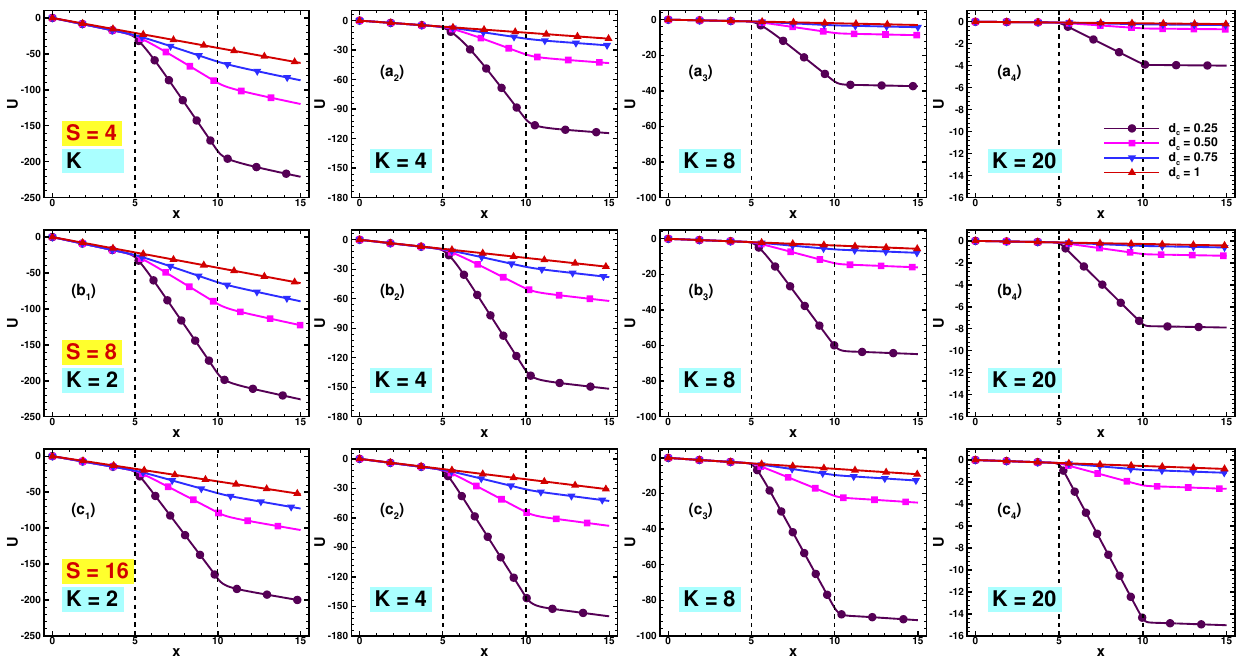}
\caption{Dimensionless total electrical potential ($U$) distribution on the centreline of charged microfluidic device as a function of dimensionless governing parameters ($K$, $S$ and $d_\text{c}$).}
\label{fig:2}
\end{figure} 
\begin{figure}[h]
\centering\includegraphics[width=0.9\linewidth]{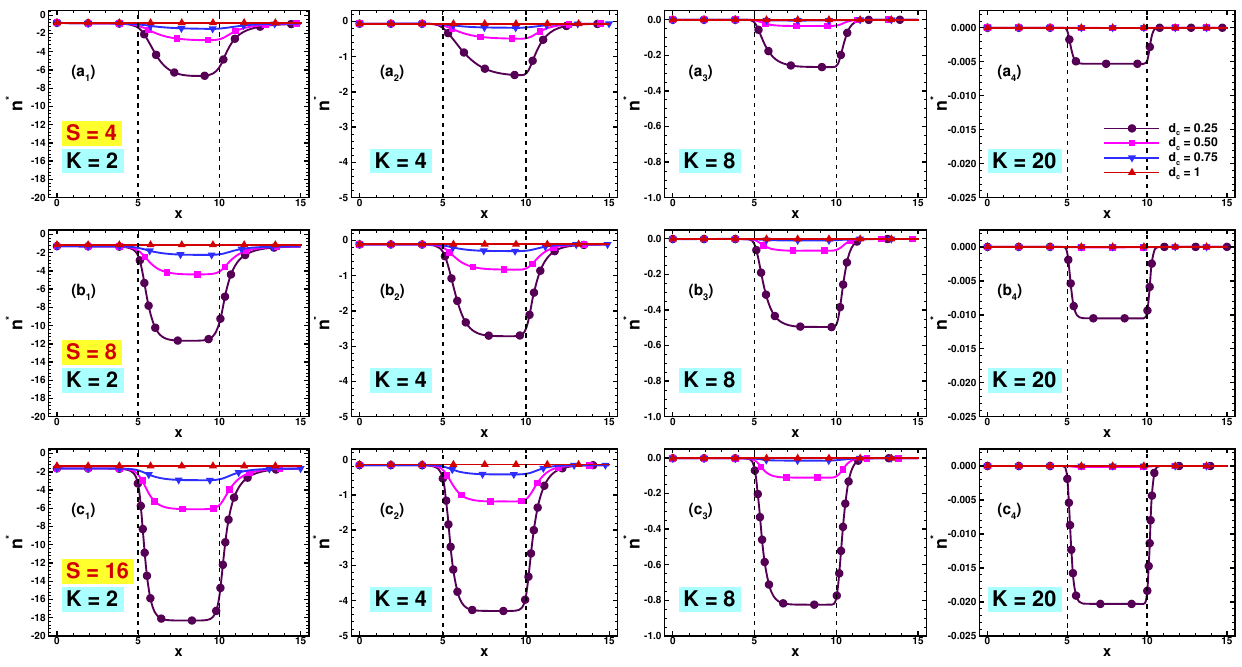}
\caption{Dimensionless excess charge ($n^\ast$) distribution on the centreline of charged microfluidic device as a function of dimensionless governing parameters ($K$, $S$ and $d_\text{c}$).}
\label{fig:3}
\end{figure} 
\begin{figure}[htbp]
\centering\includegraphics[width=1\linewidth]{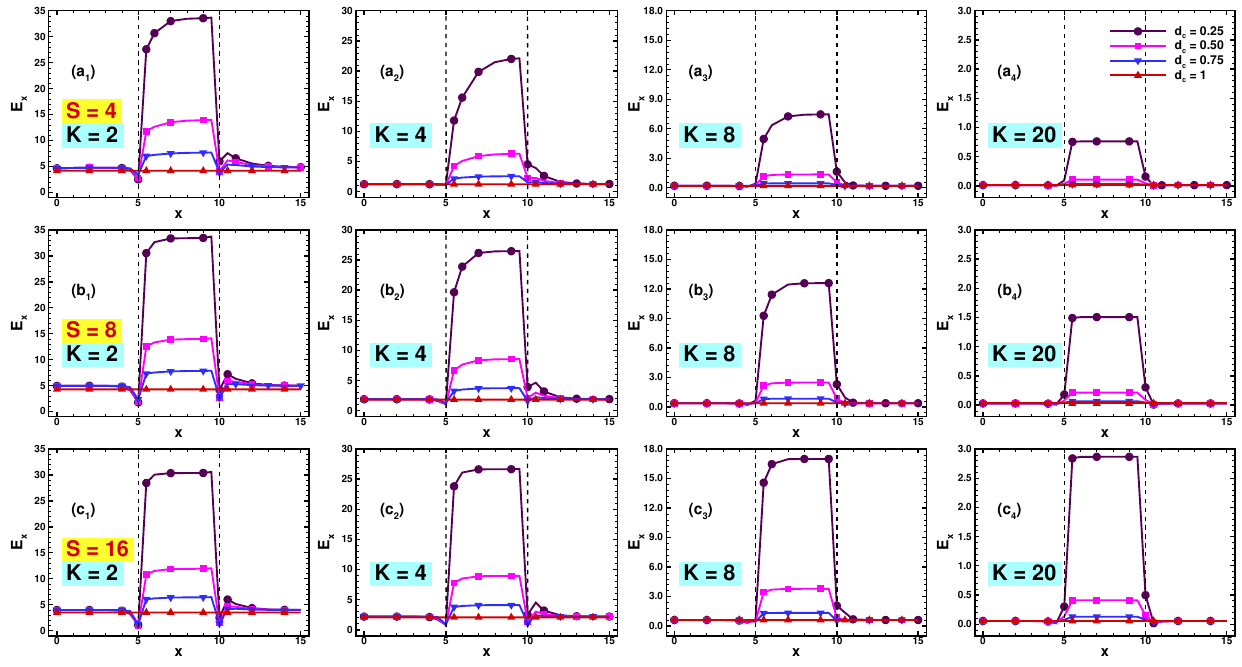}
\caption{Dimensionless induced electric field strength ($E_\text{x}$) distribution on the centreline of charged microfluidic device as a function of dimensionless governing parameters ($K$, $S$ and $d_\text{c}$).}
\label{fig:4a}
\end{figure} 
\begin{figure}[htbp]
\centering\includegraphics[width=1\linewidth]{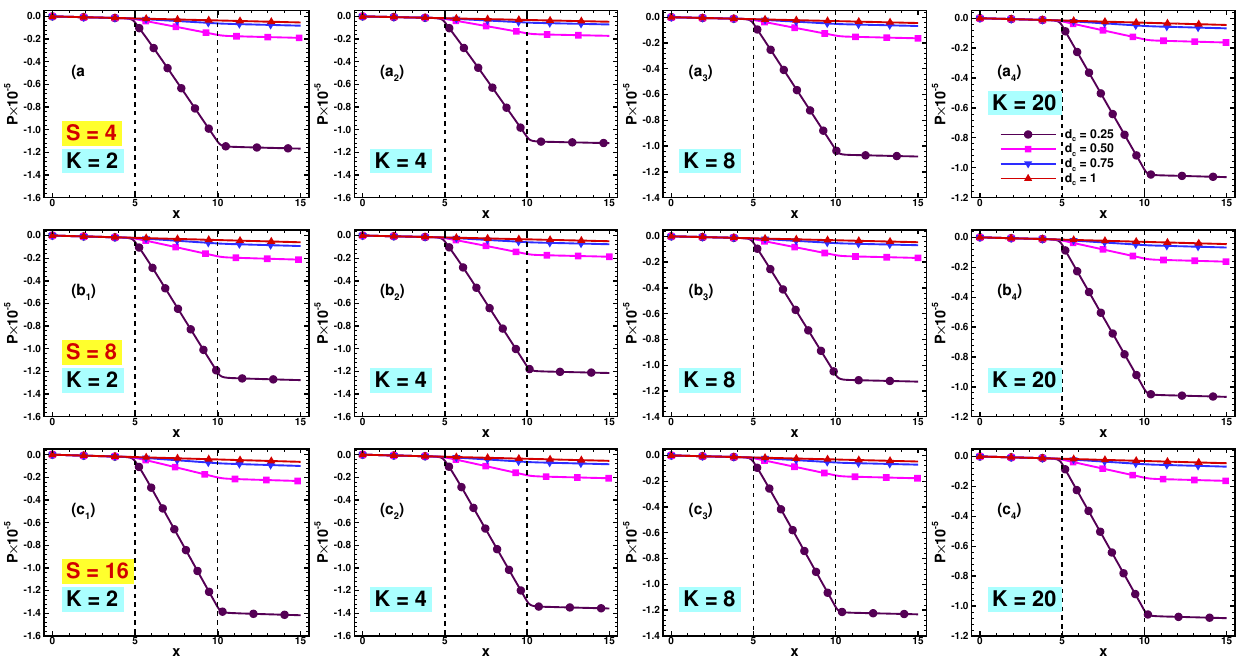}
\caption{Dimensionless pressure ($P$) distribution on the centreline of charged microfluidic device as a function of dimensionless governing parameters ($K$, $S$ and $d_\text{c}$).}
\label{fig:5}
\end{figure} 
%%
%%%
\clearpage
%--------------- Bibliography
%
%\begin{thebibliography}{0000}
%\bibliographystyle{plainnat}
%\bibliographystyle{elsarticle/elsarticle-harv}\biboptions{authoryear}
\bibliography{references}
%
% Bibliographic references with the natbib package:
% Parenthetical: \citep{Bai92} produces (Bailyn 1992).
% Textual: \citet{Bai95} produces Bailyn et al. (1995).
% An affix and part of a reference:
%   \citep[e.g.][Ch. 2]{Bar76}
%   produces (e.g. Barnes et al. 1976, Ch. 2).
% 
% \bibitem[Names(Year)]{label} or \bibitem[Names(Year)Long names]{label}.
% (\harvarditem{Name}{Year}{label} is also supported.)
%\bibitem[]{}
%
%\input{references.tex}
%
%\end{thebibliography}
%
%\addcontentsline{toc}{section}{References} 
%\clearpage\listoftables
%
%\clearpage\listoffigures
%
%\clearpage
%
%\input{tables.tex}
%
%\clearpage
%
%\renewcommand{\thesubfigure}{(\roman{subfigure})}
%
%\input{figures.tex}
%
\end{document}

%% file: nomenclature.tex
\fontsize{10}{10pt}\selectfont
 \nomenclature[g0]{\textit{Greek letters}}{}
 \nomenclature[d0]{\textit{Dimensionless groups}}{}
 \nomenclature[s0]{\textit{Subscripts and Superscripts}}{}
 \nomenclature[z0]{\textit{Abbreviations}}{}
%
%%%%%%%%%%% Abbreviations
\nomenclature[zcfd]{CFD}{computational fluid dynamics}
\nomenclature[zedl]{EDL}{electrical double layer}
\nomenclature[zevc]{EV}{electroviscous}
\nomenclature[zeve]{EVCV}{electroviscous and contraction variation}
\nomenclature[zevf]{EVF}{electroviscous flow}
\nomenclature[zfem]{FEM}{finite element method}
\nomenclature[zfep]{PDF}{pressure-driven flow}
%\nomenclature[zfvm]{FVM}{finite volume method}
%\nomenclature[zpdes]{PDEs}{partial differential equations}
%\nomenclature[zsaes]{SAEs}{simultaneous algebraic equations}
%
%%%%%%%%%%% List of Symbols
%\nomenclature[aB]{$B$}{charge-dependent slip length (\eqn\ref{eq:26}), --}
%\nomenclature[ab]{$b$}{charge-dependent slip length (\eqn\ref{eq:20}), m}
%\nomenclature[aB0]{$B_\text{0}$}{slip length (\eqn\ref{eq:26}), --}
%\nomenclature[ab0]{$b_\text{0}$}{slip length (\eqn\ref{eq:20}), m}
\nomenclature[aD]{$\mathcal{D}$}{diffusivity of the positive and negative ions, assumed equal ($\mathcal{D}_{+}=\mathcal{D}_{-}=\mathcal{D}$), m$^2$/s}
%\nomenclature[ad]{$d$}{equilibrium distance of Lennard-Jones potential ($=0.4\times 10^{-9}$,  \eqn\ref{eq:20}), m}
\nomenclature[adc]{$d_{\text{c}}$}{contraction ratio ($=W_{\text{c}}/W$, \eqn\ref{eq:12}), --}
\nomenclature[aDj]{$\mathcal{D}_{j}$}{diffusivity of the ions of type j, m$^2$/s}
\nomenclature[ae]{$e$}{elementary charge of a proton ($=1.602176634\times 10^{-19}$), C or A.s}
\nomenclature[aE]{$E_{\text{x}}$}{induced electric field strength, V/m or --}
\nomenclature[afj]{$\mathbf{f_\text{j}}$}{flux density of the ions of type j (\eqn\ref{eq:9}), 1/(m$^2$.s)}
\nomenclature[aIc]{$I_{\text{c}}$}{conduction current density (\eqn\ref{eq:7}), A/m$^2$ or --}
\nomenclature[aId]{$I_{\text{d}}$}{diffusion current density (\eqn\ref{eq:7}), A/m$^2$ or --}
\nomenclature[aIs]{$I_{\text{s}}$}{streaming current density (\eqn\ref{eq:7}), A/m$^2$ or --}
\nomenclature[akB]{$k_{\text{B}}$}{Boltzmann constant ($=1.380649\times 10^{-23}$), J/K}
%\nomenclature[alB]{$l_\text{B}$}{Bjerrum length ($=0.7\times 10^{-9}$, \eqn\ref{eq:20}), m}
\nomenclature[aLc]{$L_{\text{c}}$}{length of contraction section, m or --}
\nomenclature[aLc*]{$L^\ast_{\text{c}}$}{generalized length of contraction section, m or --}
\nomenclature[aLd]{$L_{\text{d}}$}{length of downstream outlet section, m or --}
\nomenclature[aLu]{$L_{\text{u}}$}{length of upstream inlet section, m or --}
\nomenclature[an+]{$n_{+}$}{local number density of positive ions (\eqn\ref{eq:6}), 1/m$^3$ or --}
\nomenclature[an-]{$n_{-}$}{local number density of negative ions (\eqn\ref{eq:6}), 1/m$^3$ or --}
\nomenclature[an0]{$n_{0}$}{bulk density of the ions of type j, 1/m$^3$}
\nomenclature[anj]{$n_{j}$}{local number density of the ions of type j, 1/m$^3$}
\nomenclature[ans]{$n^*$}{excess charge ($=n_{+}-n_{-}$), 1/m$^3$ or --}
%\nomenclature[ant]{$n^{**}$}{normalized excess charge, 1/m$^3$ or --}
\nomenclature[aP]{$P$}{pressure, Pa or --}
\nomenclature[aT]{$T$}{temperature, K}
\nomenclature[aU]{$U$}{total electrical potential, V or --}
\nomenclature[aV]{$\mathbf{V}$}{velocity vector, m/s or --}
\nomenclature[aVa]{$\overline{V}$}{average velocity of the fluid at the inlet, m/s}
\nomenclature[aVx]{$V_x$}{x-component of the velocity, m/s or --}
\nomenclature[aVy]{$V_y$}{y-component of the velocity, m/s or --}
\nomenclature[aW]{$W$}{cross-sectional width of inlet and outlet sections, m}
\nomenclature[aWc]{$W_{\text{c}}$}{cross-sectional width of contraction section, m}
\nomenclature[ax]{$x$}{streamwise coordinate, --}
\nomenclature[ay]{$y$}{transverse coordinate, --}
\nomenclature[aY]{$Y$}{electroviscous correction factor (\eqns\ref{eq:27}, \ref{eq:32}, and \ref{eq:28}), --}
\nomenclature[azj]{$z_{j}$}{valency of the ions of type j, assumed equal ($z_{+}=z_{-}=z$), --}
%
%------- {Greek symbols}
%
\nomenclature[gdP]{$\Delta P$}{pressure drop (\eqns\ref{eq:38} and \ref{eq:40}), --}
\nomenclature[geps0]{$\varepsilon_{\text{0}}$}{permittivity of free space (i.e. vaccum), F/m or C/(V.m)}
\nomenclature[gepsr]{$\varepsilon_{\text{r}}$}{dielectric constant (or absolute permittivity or relative permittivity) of the electrolyte liquid, --}
\nomenclature[glambdad]{$\lambda_{\text{D}}$}{Debye length $\left(=\sqrt{\frac{\varepsilon_{\text{0}}\varepsilon_{\text{r}} k_{\text{b}}T}{z^2e^2n_{\text{0}}}}\right)$, m}
\nomenclature[gmu]{$\mu$}{viscosity, Pa.s}
\nomenclature[gmueff]{$\mu_\text{eff}$}{effective or apparent viscosity, Pa.s}
\nomenclature[gpsi]{$\psi$}{EDL potential, V or --}
\nomenclature[grho]{$\rho$}{density of fluid, kg/m$^3$}
\nomenclature[grhoe]{$\rho_{\text{e}}$}{charge density of liquid, C/m$^3$}
\nomenclature[gsigmab]{$\sigma$}{surface charge density, C/m$^2$}
%\nomenclature[gsigmae]{$\sigma_{\text{e}}$}{electrical conductivity of an electrolyte solution (\eqn\ref{eq:16}), A/(V.m)}
%\nomenclature[gsigma]{$\sigma_\text{t}$}{top wall surface charge density, C/m$^2$}
%
%------- {Dimensionless group}
%
\nomenclature[dbeta]{$\mathit{\beta}$}{liquid parameter (\eqn\ref{eq:1}), --}
\nomenclature[dK]{$\mathit{K}$}{inverse Debye length (\eqn\ref{eq:1}), --}
\nomenclature[dPe]{$Pe$}{Peclet number ($={Re}~\mathit{Sc}$) (\eqn\ref{eq:1}), --}
\nomenclature[dRe]{$Re$}{Reynolds number (\eqn\ref{eq:1}), --}
\nomenclature[dSa]{$S$}{surface charge density (\eqn\ref{eq:11}), --}
\nomenclature[dSc]{$\mathit{Sc}$}{Schmidt number (\eqn\ref{eq:1}), --}
%\nomenclature[dSb]{$\mathit{S_\text{r}}$}{surface charge ratio (\eqn\ref{eq:12}), --}
%\nomenclature[dS]{$\mathit{S_\text{t}}$}{top wall surface charge density (\eqn\ref{eq:11}), --}
%
%%%%%%%%%%% subscripts and superscripts
\nomenclature[sz]{$0$}{without electroviscous effects}
\nomenclature[szz]{$00$}{without electroviscous effects and fixed contraction ($d_\text{c}=0.25$)}
%\nomenclature[scc]{$ac$}{asymmetrically charged}
\nomenclature[sc]{$c$}{contraction}
\nomenclature[sd]{$d$}{downstream}
\nomenclature[se]{$e$}{extra or excess}
\nomenclature[sm]{$m$}{mathematical}
\nomenclature[ss]{$s$}{statistical}
\nomenclature[su]{$u$}{upstream}
%\nomenclature[sN]{$N$}{normalized values}
%
%\printnomenclature